\documentclass[fleqn,usenatbib]{mnras}
\usepackage[T1]{fontenc}
\DeclareRobustCommand{\VAN}[3]{#2}
\let\VANthebibliography\thebibliography
\def\thebibliography{\DeclareRobustCommand{\VAN}[3]{##3}\VANthebibliography}
\usepackage{multirow}
\usepackage{caption}
\usepackage{subcaption}
\usepackage{float}
\usepackage{placeins}
\usepackage{comment}
\usepackage{tablefootnote}

\usepackage{graphicx}	
\usepackage{amsmath}	
\usepackage{amssymb}
\usepackage{caption}

\usepackage{scalerel,tikz}
\usetikzlibrary{svg.path}
\definecolor{orcidlogocol}{HTML}{A6CE39}
\tikzset{orcidlogo/.pic={
 \fill[orcidlogocol] svg{M256,128c0,70.7-57.3,128-128,128C57.3,256,0,198.7,0,128C0,57.3,57.3,0,128,0C198.7,0,256,57.3,256,128z};
 \fill[white] svg{M86.3,186.2H70.9V79.1h15.4v48.4V186.2z}
 svg{M108.9,79.1h41.6c39.6,0,57,28.3,57,53.6c0,27.5-21.5,53.6-56.8,53.6h-41.8V79.1z M124.3,172.4h24.5c34.9,0,42.9-26.5,42.9-39.7c0-21.5-13.7-39.7-43.7-39.7h-23.7V172.4z}
 svg{M88.7,56.8c0,5.5-4.5,10.1-10.1,10.1c-5.6,0-10.1-4.6-10.1-10.1c0-5.6,4.5-10.1,10.1-10.1C84.2,46.7,88.7,51.3,88.7,56.8z};
}}
\newcommand\orcidicon[1]{\href{https://orcid.org/#1}{\mbox{\scalerel*{
\begin{tikzpicture}[yscale=-1,transform shape]
\pic{orcidlogo};
\end{tikzpicture}
}{|}}}}

\title[Role of turbulence in z$\sim$3 SMGs]{Investigating the role of turbulence in the interstellar medium in $z\sim3$ dusty star-forming galaxies using kpc-resolution ALMA dust and gas maps}
\author[B. A. Westoby et al.]{B. A. Westoby,$^{\orcidicon{0009-0004-3732-6394}\,1}$\thanks{E-mail: westoby@strw.leidenuniv.nl},
J. A. Hodge,$^{\orcidicon{0000-0001-6586-8845}\,1}$\thanks{E-mail: hodge@strw.leidenuniv.nl}, 
P. Sharda,$^{\orcidicon{0000-0003-3347-7094}\,1}$,
P. E. Mancera Piña $^{\orcidicon{0000-0001-5175-939X}\,1}$, 
M. Rybak $^{\orcidicon{0000-0002-1383-0746}\,1,2,3}$,
\newauthor
E. da Cunha $^{\orcidicon{0000-0001-9759-4797}\,4,5}$,
J. Li $^{\orcidicon{0000-0002-8184-5229}\,4}$,
I. Smail $^{\orcidicon{0000-0003-3037-257X}\,6}$,
A. M. Swinbank $^{\orcidicon{0000-0003-1192-5837}\,6}$,
A. Battisti $^{\orcidicon{0000-0003-4569-2285}\,4,5}$,
L. A. Boogaard $^{\orcidicon{0000-0002-3952-8588}\,1}$,
\newauthor
W. N. Brandt $^{\orcidicon{0000-0002-0167-2453}\,7,8,9}$,
G. Calistro Rivera $^{\orcidicon{0000-0003-0085-6346}\,10,11}$,
C.-C. Chen$^{\orcidicon{0000-0002-3805-0789}\,12,13}$,
P. Cox$^{\orcidicon{0000-0003-2027-8221}\,14}$,
M. Cracraft$^{\orcidicon{0000-0002-3805-0789}\,15}$,
\newauthor
H. Dannerbauer$^{\orcidicon{0000-0001-7147-3575}\,16}$,
R. Decarli$^{\orcidicon{0000-0002-2662-8803}\,17}$,
T. R. Greve$^{\orcidicon{0000-0002-2554-1837}\,18,19,20}$,
S. Kendrew$^{\orcidicon{0000-0002-7612-0469}\,21}$,
K. Knudsen$^{\orcidicon{0000-0002-7821-8873}\,22}$,
\newauthor
C.-L. Liao$^{\orcidicon{0000-0002-5247-6639}\,1}$,
J. van Marrewijk$^{\orcidicon{0000-0001-9830-3103}\,1}$,
O. Nayak$^{\orcidicon{0000-0001-6576-6339}\,23}$,
M. Neeleman$^{\orcidicon{0000-0002-9838-8191}\,24}$,
L. E. Rowland$^{\orcidicon{0009-0009-2671-4160}\,1}$,
\newauthor
E. Schinnerer $^{\orcidicon{0000-0002-3933-7677}\,25}$,
F. Walter $^{\orcidicon{0000-0003-4793-7880}\,25}$,
J. L. Wardlow $^{\orcidicon{0000-0003-2376-8971}\,26}$,
A. Weiss$^{\orcidicon{0000-0003-4678-3939}\,27}$,
P. van der Werf $^{\orcidicon{0000-0001-5434-5942}\,1}$
\\
$^{1}$Leiden Observatory, Leiden University, P.O. Box 9513, 2300 RA Leiden, The Netherlands \\ $^{2}$Faculty of Electrical Engineering, Mathematics and Computer Science, Delft University of Technology, Mekelweg 4, \\ 2628 CD Delft, The Netherlands \\ 
$^{3}$SRON -- Netherlands Institute for Space Research, Niels Bohrweg~4, 2333 CA Leiden, The Netherlands \\
$^{4}$International Centre for Radio Astronomy Research, University of Western Australia, 35 Stirling Hwy, Crawley 26WA 6009, Australia \\
$^{5}$Research School of Astronomy and Astrophysics, Australian National University, Cotter Road, Weston Creek, ACT 2611, Australia \\
$^{6}$Centre for Extragalactic Astronomy, Department of Physics, Durham University, South Road, Durham DH1 3LE, UK \\ 
$^{7}$Department of Astronomy and Astrophysics, 525 Davey Lab, The Pennsylvania State University, University Park, PA 16802, USA \\
$^{8}$Institute for Gravitation and the Cosmos, The Pennsylvania State University, University Park, PA 16802, USA \\
$^{9}$Department of Physics, 104 Davey Laboratory, The Pennsylvania State University, University Park, PA 16802, USA \\
$^{10}$ German Aerospace Center (DLR), Institute of Communications and Navigation, Wessling, Germany \\ 
$^{11}$ European Southern Observatory (ESO), Karl-Schwarzschild-Straße 2, 85748 Garching bei M{\"u}nchen, Germany \\
$^{12}$Academia Sinica Institute of Astronomy and Astrophysics (ASIAA), 
11F of Astronomy-Mathematics Building, 
AS/NTU, No. 1, Sec. 4, \\ Roosevelt Road, Taipei 106319, Taiwan \\
$^{13}$East Asian Observatory, 660 N. A'ohoku Pl., Hilo, HI 96720, USA \\
$^{14}$ Sorbonne Universit{\'e}, UPMC Universit{\'e} Paris 6 and CNRS, UMR 7095, Institut d'Astrophysique de Paris, 98bis boulevard Arago, \\ 75014 Paris, France \\
$^{15}$ Space Telescope Science Institute, 3700 San Martin Drive, Baltimore MD 21218, USA \\
$^{16}$Instituto de Astrof\'isica de Canarias, V\'ia L\'actea, 39020 La Laguna (Tenerife), Spain \\
$^{17}$ INAF -- Osservatorio di Astrofisica e Scienza dello Spazio di Bologna, via Gobetti 93/3, I-40129, Bologna, Italy \\
$^{18}$Cosmic Dawn Center (DAWN), Denmark \\
$^{19}$DTU Space, Technical University of Denmark, Elektrovej, Building 328, 2800, Kgs. Lyngby, Denmark \\
$^{20}$Dept. of Physics and Astronomy, University College London, Gower Street, London WC1E 6BT, UK \\
$^{21}$European Space Agency (ESA), ESA Office, Space Telescope Science Institute, 3700 San Martin Drive, Baltimore, MD 21218, USA \\
$^{22}$Department of Space, Earth and Environment, Chalmers University of Technology, SE-412 96 Gothenburg, Sweden \\
$^{23}$ NASA Goddard Space Flight Center, 8800 Greenbelt Road, Greenbelt, MD, USA \\
$^{24}$ National Radio Astronomy Observatory, 520 Edgemont Road, Charlottesville, VA 22903, USA\\
$^{25}$ Max--Planck Institut f\"ur Astronomie, K\"onigstuhl 17, 69117, Heidelberg, Germany\\
$^{26}$Department of Physics, Lancaster University, Lancaster, LA1 4YB, UK \\
$^{27}$Max-Planck-Institut f\"ur Radioastronomie, Auf dem H\"ugel 69, D-53121 Bonn, Germany}

 \date{Accepted XXX. Received YYY; in original form \today}
\pubyear{\the\year{}}
\begin{document}
\label{firstpage}
\pagerange{\pageref{firstpage}--\pageref{lastpage}}
\maketitle

\clearpage

\begin{abstract}
We present ALMA high-resolution ($\sim 0.25\arcsec /2\,\rm{kpc}$) CO(5--4) and CO(4--3) observations of three $z\sim 3$ submillimetre-selected dusty galaxies from the ALESS survey. 
These data complement existing [sub]-kpc scale ALMA 870$\mu$m continuum imaging and JWST NIRCam and MIRI imaging from the ALESS-JWST program, allowing us to trace the molecular gas, dust-obscured star formation, and stellar populations on similar spatial scales. We spectroscopically confirm that two of the sources lie at the same redshift and are likely interacting.
We find that the molecular-gas distribution broadly follows the dusty star-forming structures seen in the 870$\mu$m dust continuum imaging, but that the gas reservoirs are significantly more extended than the dust emission with a spatial extent comparable to the rest-frame near-infrared stellar emission.
By modeling the kinematics for the two highest signal-to-noise sources, 
we find that the galaxies are well-fit by rotating disc models with high ratios of ordered to random motion ($V_{\rm{max}}\,/\,\overline{\sigma}=5\pm1$ and $6\pm1$), 
although smaller-scale kinematic deviations cannot be ruled out at the current sensitivity and spatial resolution.
Finally, utilizing the high-resolution 870$\mu$m dust continuum and CO data, we investigate star-formation scaling relations on kpc-scales in these high-redshift galaxies. 
Assuming a constant CO-to-H$_{2}$ conversion factor and excitation ratio, we find that the data are offset from theoretical star-formation relation predictions that do not take turbulence into account, but consistent with gravo-turbulent models, thereby suggesting that turbulence plays a central role in regulating star formation at high redshift. 
\end{abstract}

\section{Introduction}
The advent of the James Webb Space Telescope (JWST) has provided new insights into dusty star formation in the early Universe. A significant fraction of this dust-obscured star formation is thought to be hosted by submillimetre galaxies \citep[SMGs,][]{dudzeviciute_alma_2020}, which were originally discovered with the SCUBA instrument mounted on the James Clerk Maxwell telescope \citep{smail_deep_1997,barger_submillimetre-wavelength_1998,hughes_high-redshift_1998,eales_canada-uk_1999}. Now, the near- and mid-infrared capabilities of JWST are uncovering the previously dust-obscured stellar populations in these sources, revealing their underlying rest-frame optical morphologies on resolved scales for the first time \citep[e.g.,][]{chen_jwst_2022,gillman_structure_2024,le_bail_jwstceers_2024,akins_jwstalma_2025,bodansky_jwstalma_2025,hodge_aless-jwst_2025,ikeda_formation_2025,mckinney_scubadive_2025,uematsuALMASCUBA2COSMOS2025a,boogaard_resolving_2026}.
These studies add to the work done in the last decade on directly observing the dust  continuum emission itself in SMGs with ALMA on similar scales \citep[see the review by][]{hodge_high-redshift_2020}, with the latest direct comparisons between the two components, definitively demonstrating that the galaxies are primarily red due to dust and not stellar age \citep{gillman_structure_2024,hodge_aless-jwst_2025}. 

Despite this progress, there are only a handful of studies that have looked at the molecular gas content in these sources on similar ($\sim$kpc) scales \citep[e.g.,][]{swinbank_ism_2011, hodge_evidence_2012,calistro_rivera_resolving_2018}. This is partly because many SMGs still lack spectroscopic redshifts due to being very faint and/or dust-obscured. This has recently improved thanks to dedicated spectral line scans with ALMA, enabling both redshift confirmation and investigation of the (unresolved) molecular gas properties via the detected CO lines \citep[e.g.,][]{birkin_almanoema_2021,liao_alma_2024}.\\

One motivation for resolving the gas in SMGs is to use kinematic characterization to provide a better understanding of the nature of these sources. High-fidelity ALMA 870$\mu$m continuum maps revealing the morphology of the dust continuum in distant SMGs have shown that a fraction of their emission can arise from clumpy substructure \citep{hodge_alma_2019,ikeda_formation_2025}. This substructure could provide clues on how SMGs achieve and sustain high star-formation rates (SFRs). Evidence of ordered rotation of SMGs, as found in some sources \citep[e.g.,][]{swinbank_ism_2011, calistro_rivera_resolving_2018, amvrosiadis_kinematics_2025}, would suggest the substructures are disc-like features and most likely induced by a minor merger or tidal disturbance. These dust continuum structures may be coherent, indicating bars, star-forming rings or spiral arms \citep[][]{hodge_alma_2019, gullberg_alma_2019, tsukui_spiral_2021,tsukui_detecting_2023}. Such non-axisymmetric features would increase the galaxy's star formation efficiency and allow for net angular momentum loss by forcing the gas streams to cross and shock \citep{hopkins_analytic_2011}.
However, kinematic evidence to support this is limited \citep[e.g.,][]{umehata_adf22-web_2025}. Alternatively, perturbed velocity fields, as found in some other sources \citep[e.g.,][]{tadaki_noncorotating_2020,frias_castillo_kiloparsec-scale_2022}, would instead suggest that the clumpy morphologies may be due to late-stage minor/major mergers. However, so far there has been a lack of joint analyses of sources with both high-resolution CO imaging and confirmed dusty substructures.

Another motivation for looking at the resolved gas distributions in SMGs is to constrain the relative importance of ordered-to-random motion. There have been an increasing number of studies that find evidence for dynamically cold ($V_{\rm{rot}}\,/\,\sigma> 2$) rotating discs at $z > 3$ when looking at emission lines such as CO or [C{\sc ii}] \citep[e.g.][]{hodge_evidence_2012,
lelli_massive_2021,rizzo_dynamical_2021,
roman-oliveira_regular_2023,
rowlandREBELS25DiscoveryDynamically2024a}.
This may be due to the tracer used, with studies using H$\alpha$ kinematics showing higher values of velocity dispersion than that deduced from cold gas tracers \citep{alaghband-zadeh_integral_2012,rizzo_alma-alpaka_2024,birkin_kaoss_2024,danhaiveDawnDiscsUnveiling2025}.
However, we would like to investigate this for classically-selected SMGs, where interaction-induced turbulence may be important, 
leading to higher gas velocity dispersions.\\

Resolved gas properties of SMGs can also help us better understand how turbulence influences star formation in these galaxies. 
A number of relations have been theorised that link the SFR with the molecular gas mass \citep[e.g.][]{kennicutt_global_1998,de_los_reyes_revisiting_2019,pessa_star_2021,wang_3_2022}, free-fall time \citep[e.g.][]{krumholz_universal_2012} and turbulence \citep[e.g.][]{salim_universal_2015}. Their consistency with observations has been investigated with both and high redshift sources \citep[e.g.][]{kennicutt_global_1998,heiderman_star_2010,gutermuth_correlation_2011,hodge_kiloparsec-scale_2015,jameson_relationship_2016,chen_spatially_2017,bethermin_alma-alpine_2023}.
However, at high redshift, the availability of resolved maps of both star formation and molecular gas -- providing insight into the turbulence in the interstellar medium (ISM) -- is still a relatively rare combination in unlensed targets, and the resulting star formation `laws' have thus only been explored in a select few exceptionally bright and/or lensed SMGs \citep[e.g.][]{sharda_testing_2018,sharda_testing_2019}. 
High-resolution ALMA observations of more representative SMGs provide a unique opportunity to study the complexity of star formation at sub-galactic scales in the early Universe, and to directly probe the role of turbulence in regulating star formation under typical high-redshift conditions. \\

Here we present new ALMA CO observations of three $z \sim 3$ SMGs from the ALMA follow-up of the LABOCA Extended Chandra Deep Field South Survey
\citep[ALESS,][]{hodge_alma_2013, karim_alma_2013}. These data are complemented by existing high-resolution 870$\mu$m continuum imaging from \cite{hodge_alma_2019} and multi-band NIRCam imaging 
\citep[ALESS-JWST,][]{hodge_aless-jwst_2025}. In this work, we use a standard $\Lambda$CDM cosmology from \cite{planck_collaboration_planck_2020}. 

\section{Observations and Data Reduction}

\subsection{ALMA CO(5--4) and CO(4--3) Observations}
\label{sec:sample}
We present new CO observations for three SMGs (ALESS3.1, ALESS3.1-comp and ALESS9.1). These targets are selected from the ALESS survey \citep{hodge_alma_2013} based on the existence of high-resolution $870\,\rm{\mu m}$ follow-up imaging \citep[][]{hodge_kiloparsec-scale_2016,hodge_alma_2019} and confirmed CO line detections \citep{birkin_almanoema_2021}, with the exception of ALESS3.1-comp which was serendipitously detected in CO emission in this work. Information about the sources is shown in Table~\ref{tab:targetinfo}, with the positions from \cite{hodge_aless-jwst_2025}, and redshifts from \cite{birkin_almanoema_2021} for all sources except ALESS3.1-comp (which was derived from the CO data presented in this work).
The stellar masses for all sources were derived using integrated SED fitting including the ALMA and JWST NIRCam (and MIRI) imaging (\citealt{li_aless--jwst_2026}; and see Section \ref{NIRCam_pmap} for further details). 
The median continuum flux density at $870\mu\rm{m}$, $S_{\rm 870} \sim 8.3\,\rm{mJy}$, is higher than the median of the 13 galaxies covered in the ALESS-JWST program from \citet{hodge_aless-jwst_2025}. 
The stellar masses and SFRs of the sources are broadly representative of the parent ALESS sample of 99 SMGs in \cite{da_cunha_alma_2015} which have a median log($M_*/\rm{M_\odot}$)=$10.95^{+0.6}_{-0.8}$ and log(SFR/$\rm{M_\odot yr}^{-1}$)=$2.45^{+0.4}_{-0.5}$. The median redshift $z=3.5$ of the sources is higher than the parent sample median redshift of $z=2.7\pm0.1$ \citep{da_cunha_alma_2015}.
All of the SMGs targeted are consistent with lying on the star formation main sequence from \cite{speagle_highly_2014} \citep[see][]{li_aless--jwst_2026}.

The CO emission line data for this work were observed in May 2023 for project 2022.1.00955.S [PI: Hodge]. We observed CO(5--4) emission in ALESS3.1 and ALESS3.1-comp using Band 4, with a total exposure time of 160 minutes on source. The baseline lengths were 15--3143 m and the average precipitable water vapour (PWV) was 1.80 mm. We observed CO(4--3) emission in ALESS9.1 using Band 3, with a total time of 120 minutes spent on source. The baseline lengths were 27--3637 m and the average PWV was 0.86 mm. For all sources, the resolution of the spectral window covering the CO line was set to 1.953 MHz and the other spectral windows were set to 31.250 MHz. We note that a single observation was also conducted as part of this project for an additional source: ALESS17.1. As no CO line is detected for the SMG, we discuss the analysis of this source in Appendix \ref{sec:17}.

\begin{table*}
  \centering
  \caption{Basic properties of the targets. Positions and $S_{870}$ are from \protect\cite{hodge_aless-jwst_2025}. $\rm{M}_*$ and SFR are from integrated SED fitting from \protect\cite{li_aless--jwst_2026}. The ALESS3.1-comp redshift is from this work.}
  \begin{tabular}{llccccc}
    \hline
    Source ID & $z$ & R.A. (J2000) & Dec. (J2000) & $S_{870}$ (mJy) & $\rm{M}_{\rm{*}}$ [$10^{11}$ $\rm{M}_\odot$] & SFR ($\rm{M}_\odot \rm{yr}^{-1}$)\\
    \hline
    ALESS3.1   & 3.375   & 03:33:21.51 & -27:55:20.5   & 8.3 ± 0.4 & 
    $2.3^{+1.1}_{-0.7}$
    & $760^{+150}_{-110}$  \\
    ALESS3.1-comp & 3.374 & 03:33:21.43 & -27:55:25.4 & 1.07 ± 0.06 & $0.17^{+0.06}_{-0.06}$ & $60^{+40}_{-26}$    \\
    ALESS9.1   & 
    3.694
    & 03:32:11.33 & -27:52:12.0 & 8.8 ± 0.5 & $2.9^{+1.1}_{-0.9}$ & $930^{+140}_{-160}$  \\
    \hline
  \end{tabular}
  \label{tab:targetinfo}
\end{table*}

The data were calibrated using the standard ALMA pipeline with \textsc{casa} version 6.4.1.12 \citep{mcmullin_casa_2007,hunter_alma_2023}. We examined the data using the \texttt{plotms} task and synthesised images (see below). We found the data quality to be excellent, without a need for further manual flagging.
For each target, a calibrated measurement set (ms) was created using the \texttt{concat} task to combine the observations from every observing block.

To derive the CO properties, the \texttt{uvcontsub} task was used to remove the continuum from the calibrated ms file, using a linear fit to the spectral windows without the line and the channels in the spectral window containing the line that are outside the line. 
The \texttt{tclean} task was used in cube mode with natural weighting and a cell size of 0.04$''$ to produce a "dirty" cube in order to measure the rms noise per beam per channel. The cube is also converted into velocity units using the reference frequency calculated based on the redshift in Table~\ref{tab:targetinfo} and the rest frequency of the expected CO emission 
($\nu_{\rm{CO}(4-3)}=461.04\rm{GHz}\,,\,\nu_{\rm{CO}(5-4)}=576.27\rm{GHz}).$
The data are channel averaged to channels of $\sim$20 km s$^{-1}$. 
The cube is then cleaned using \texttt{tclean} in non-interactive mode with a threshold of double the rms and a one arcsecond radius circular mask around the source centres. The cleaned CO(5--4) cube for ALESS3.1 and ALESS3.1-comp has a beam size of $0.28''\times0.23''$ RMS per channel of 140 $\mu$Jy beam$^{-1}$. The cleaned CO(4--3) cube for ALESS9.1 has a beam size of $0.32''\times0.26''$ and RMS per channel of 136 $\mu$Jy beam$^{-1}$. These angular resolutions correspond to spatial scales of $\sim$2 kpc at $z\sim3$.

Furthermore, we present imaging of the continuum underlying the CO emission in the above observations. To measure this continuum, the \texttt{tclean} task was used in the multi-frequency synthesis (mfs) mode with a cell size of 0.04$''$, collapsing the off-line channels to produce a "dirty" image in order to measure the rms noise per beam. The cube is then cleaned using \texttt{tclean} with a threshold of double the rms and a one arcsecond radius circular mask around the source centers. The resulting continuum images are at 2170$\mu$m (high-resolution) 
for ALESS3.1 and ALESS3.1-comp and 3260$\mu$m for ALESS9.1 
and shown in Appendix \ref{sec:map}. \\

This work additionally uses lower-resolution CO emission line data observed in July 2021 for project 2019.1.00771.S [PI: Rybak]. This project targeted CO(4--3) emission in ALESS3.1 and ALESS3.1-comp using Band 3, with a total time of 46 minutes on source. The baseline lengths were 28--3638m and the average PWV was 0.19mm. Given the lower angular resolution ($\sim$0.35$''$), the majority of this work focuses on the higher-resolution data from project 2022.1.00955.S above unless otherwise specified. 

\subsection{870$\mu$m Continuum data}
\label{sec:continuumdata}

The ALMA 870$\mu$m continuum imaging used for this work was taken as part of programs 2012.1.00307.S and 2016.1.00048.S and first presented in \cite{hodge_kiloparsec-scale_2016, hodge_alma_2019}. For the targets in this work, the average angular resolution for the naturally weighted maps is 0.10$''$$\times$0.07$''$ and the average RMS noise is 19 $\mu$Jy beam$^{-1}$. The 870$\mu$m images, despite their high angular resolution, were shown to recover all of the emission from these targets \citep{hodge_alma_2019} when suitably tapered.

\subsection{JWST NIRCam images and parameter maps}
\label{NIRCam_pmap}
The JWST NIRCam data used for this work were observed between 2022 September and December for the GO Cycle 1 program PID 2516 (PIs: Hodge and da Cunha) and first presented in \cite{hodge_aless-jwst_2025}. Our targets are all imaged in the F200W, F356W and F444W filters and aligned using the Gaia DR3 catalogue (\citealt{gaia_collaboration_gaia_2023}). The absolute astrometric accuracy was estimated to be within 35 mas \citep{hodge_aless-jwst_2025}. Resolved and integrated SED fitting using \textsc{magphys} \citep{da_cunha_simple_2008, da_cunha_alma_2015, battisti_strength_2020} that includes these data, high-resolution 870$\mu$m ALMA data (Section \ref{sec:continuumdata}) and MIRI F770W imaging (also from PID 2516) is detailed in \cite{li_aless--jwst_2026}. 

\section{Results and analysis}
\subsection{Spectra and imaging}
\label{sec:SpectraAndImaging}

\begin{table*}
\caption{Results from the CO emission line imaging: CO(5--4) for ALESS3.1, ALESS3.1-comp and CO(4-3) for ALESS9.1. Flux is measured from the zeroth moment maps, FWHM is calculated from Gaussian fitting of the spectra, $\mu_{\rm{gas}}=M_{\rm{gas}}/M_{\star}$.}
\centering
    \begin{tabular}{|l|c|c|c|c|c|c|c|}
    \hline
    Source ID  & Flux [$\rm{Jy\,km\,s^{-1}}$] & FWHM [$\rm{km\,s^{-1}}$] & SNR & $L'_{\rm{CO,midJ}}$ [$10^{11}\rm{\,K\,km\,s^{-1}}pc^2$] & $\rm{M}_{\rm{gas}}$ [$10^{11}$ $\rm{M}_\odot$]  & $\mu_{\rm{gas}}$
    \\ \hline
    ALESS3.1 & 1.62 $\pm$ 0.14 & 677 ± 68 & 11.45 & 0.33 $\pm$ 0.03 & 0.90 $\pm$ 0.19 & $0.39^{+0.15}_{-0.21}$\\ 
    ALESS3.1-comp & 0.23 $\pm$ 0.05 & 187 ± 43 & 5.07 & 0.05 $\pm$ 0.01 & 0.13 $\pm$ 0.04 & $0.8^{+0.4}_{-0.4}$\\ 
    ALESS9.1 & 1.34 $\pm$ 0.12 & 630 ± 63 & 10.70 & 0.49 $\pm$ 0.04  & 1.43 $\pm$ 0.21 & $0.50^{+0.18}_{-0.20}$\\ \hline
    \end{tabular}
    \label{tab:imaging}
\end{table*}

\begin{figure}
    \centering
    \includegraphics[width=\columnwidth]
    {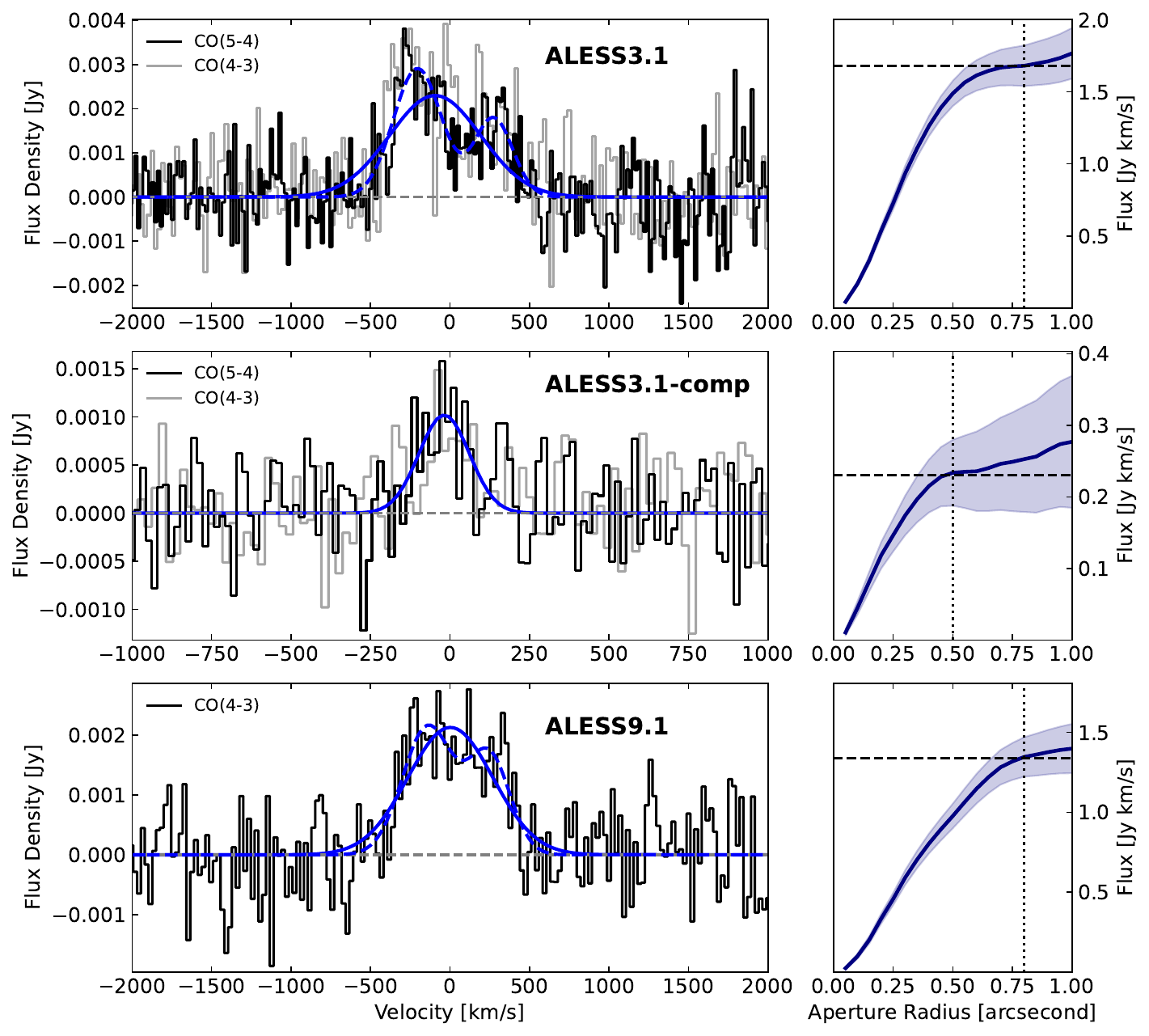}
    \caption{CO spectra and curve-of-growth analysis showing flux as a function of aperture radius. The left hand panels show the single Gaussian fits to the spectra, we also show a dashed double Gaussian for ALESS3.1 and ALESS9.1. The dotted vertical black lines in the right-hand panels are the apertures used to extract the spectra, the dashed horizontal black lines are the final fluxes extracted from the zeroth moment maps.}
    \label{fig:ALMACOG}
\end{figure}

We detect strong CO emission in all naturally-weighted image cubes. We make initial moment maps by extracting spectra from the image cubes using a circular aperture of 1$\arcsec$ radius around the source, fitting a single Gaussian to each spectrum and collapsing over $\pm\,1\times$FWHM from the central frequency. Then, we conduct a curve-of-growth analysis using \texttt{imstat} to find the optimum aperture radius from which to extract the total flux. This is a non-parametric method from which we extract fluxes by identifying the aperture radius at which the measured flux converges to a plateau \citep[e.g.][]{chen_extended_2020,frias_castillo_vla_2023}. We extract final spectra (Fig.~\ref{fig:ALMACOG}) and fluxes using that aperture: for ALESS3.1 and ALESS9.1, the aperture radius was 0.8$\arcsec$, and for ALESS3.1-comp the aperture radius was 0.5$\arcsec$. 

\begin{figure*}
    \includegraphics[width=0.8\linewidth]{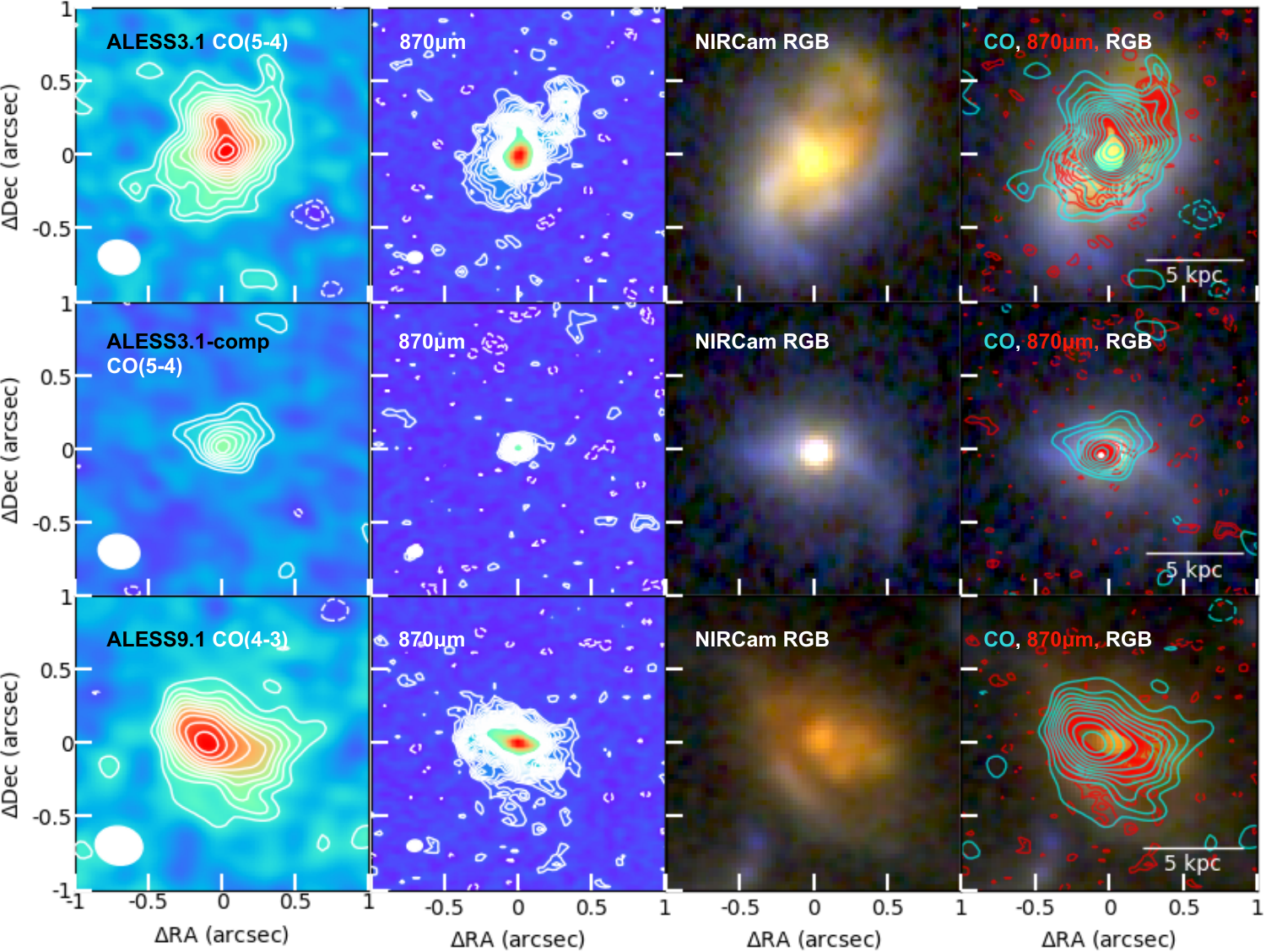}
    \caption{Naturally weighted $2''\times2''$ zeroth moment maps of the CO line emission (1st column) alongside dust continuum maps \protect\citep{hodge_alma_2019} (2nd column) and NIRCam RGB(F444W/F356W/F200W) images \protect\citep{hodge_aless-jwst_2025} (3rd column), centered on the positions in Table \ref{tab:targetinfo}. The 4th column presents an overlay of all maps. The cyan CO contours and red dust contours start at $\pm2\sigma$ with 1$\sigma$ steps, positive contours are solid and negative contours are dashed.}
    \label{fig:multi-tracer}
\end{figure*}

All spectra are fit with a single Gaussian. The image cubes are then collapsed over $\pm$1$\times$FWHM of the final spectra to produce the final zeroth moment maps (Fig.~\ref{fig:multi-tracer}) from which the fluxes\footnote{We check for the impact of the 'JvM effect' \citep{jorsater_high_1995} in our cleaned high-resolution maps. This flux-scaling issue, arising because the clean model and residual map have different units (Jy per clean beam versus Jy per dirty beam), can affect flux measurements in low S/N regions of extended emission. Following the procedure outlined in \cite{rybakCO10ImagingReveals2025}, we measure the flux densities for the sources in the $uv$-plane, which is not impacted by the 'JvM effect'. We find that the flux densities measured in the $uv$-plane are consistent with those measured in the image plane, suggesting that the impact of this effect is negligible.} are measured. The results of the imaging are shown in Table~\ref{tab:imaging}. The images produced from the lower-resolution CO data (project 2019.1.00771.S) can be found in Appendix \ref{sec:map}. 

Both ALESS3.1 and ALESS9.1 show evidence for double peaked line properties. ALESS 3.1 exhibits a double-horn profile that is significantly asymmetric. When fit with a double Gaussian, the peak CO(5-4) flux density for the high-frequency wing is $\sim60\%$ higher than the low-frequency wing and the FWHMs are $360\pm50\,\rm{km\,s^{-1}}$ and $290\pm70\,\rm{km\,s^{-1}}$ for the high and low frequency wings respectively, with the peaks separated by $\sim480\,\rm{km\,s^{-1}}$. The profile of ALESS9.1 is more symmetric. When fit with a double Gaussian, the peak CO(4--3) flux density for the high-frequency wing is only $\sim$ 20\% higher than the low-frequency wing and the FWHMs are $370\pm100\,\rm{km\,s^{-1}}$ and $310\pm100\,\rm{km\,s^{-1}}$ for the high and low frequency wings respectively, with the peaks separated by $\sim385\,\rm{km\,s^{-1}}$.

The detections of CO(5--4) and CO(4--3) in the spectra of ALESS3.1-comp allow us to determine its (previously unknown) spectroscopic redshift, finding  $z=3.374$. This suggests that there is only a 40 km s$^{-1}$ offset between it and ALESS3.1 and a separation on the sky of $\sim$ 40 kpc. 

We calculate line luminosities $L'_{\mathrm{CO,J}}$ (in $\rm{K\,km\,s^{-1}\,pc^2}$), following \cite{solomon_molecular_2005}: 
\begin{equation}
L'_{\rm{CO,J}} = 3.25\times10^7 I_{\rm{CO}} \nu_{\rm{obs}}^{-2}D_{\rm{L}}^2(1+z)^{-3}
\end{equation}
where $\nu_{\mathrm{obs}}$ is the reference frequency of the CO line in GHz and $D_{\rm{L}}$ is the luminosity distance in Mpc.
We then convert luminosity $L'_{\mathrm{CO,J}}$ into $L'_{\mathrm{CO(1-0)}}$ assuming the typical ratios from \cite{birkin_almanoema_2021}: $r_{41}=0.34\pm0.04$ and $r_{51}=0.36\pm0.07$. We find molecular gas masses $M_{\rm{gas}}$
by multiplying $L'_{\mathrm{CO(1-0)}}$ by a constant $\alpha_{\rm{CO}}$=1\,$\rm{M_\odot}\, (K\,km\,s^{-1}\,pc^2)^{-1}$ \citep[including Helium, see][]{bothwell_survey_2013,birkin_almanoema_2021,frias_castillo_kiloparsec-scale_2022,frias_castillo_comparative_2025}. 
For the two sources with [C{\sc i}] detections (ALESS3.1 and 9.1) we verify that the H$_2$ masses derived here are consistent within $\sim 1\sigma$ with those derived based on the [C{\sc i}] detections in \cite{birkin_almanoema_2021}, using equation 4 from \cite{frias_castillo_comparative_2025} with $Q_{10}=0.46$ and $X_{[\rm{CI}]}=(5.1\pm1.8)\times10^{-5}$. Furthermore, we also derive a gas mass for all three sources using integrated dust masses from \cite{li_aless--jwst_2026} and, using an SMG gas-to-dust ratio \citep[$\delta_{\rm{gdr}}/\alpha_{\rm{CO}}=63\pm7$,][]{birkin_almanoema_2021}, we again find them consistent within $\sim 1\sigma$. Using the stellar masses from Table~\ref{tab:targetinfo} and the gas masses from Table~\ref{tab:imaging}, we calculate the molecular gas to stellar mass ratio as $\mu_{\rm{gas}}=M_{\rm{gas}}/M_{\star}$. We list these values in Table~\ref{tab:imaging}.\\

Fig.~\ref{fig:multi-tracer} shows $2''\times2''$ cutouts of the naturally weighted high-resolution CO zeroth-moment maps, 870$\mu$m continuum imaging and NIRCam RGB images with red as F444W, green as F356W and blue as F200W. The galaxies are also detected and resolved in the continuum underlying the targeted CO lines, but at lower spatial resolution and S/N than with the existing 870$\mu$m imaging; the new continuum maps are shown in the appendix in Fig. \ref{fig:COcont}. 

We see that the gas traced by the CO observations is approximately co-located with the 870$\mu$m dust continuum and roughly follows its morphology, as might be expected. In particular, we see in ALESS3.1 that the arm/tidal tail feature visible in the 870$\mu$m dust continuum in the northwestern region is also detected in the CO(5--4) emission. We also observe that the CO map has an extension north of the nucleus that follows a similar extension in the 870$\mu$m dust continuum. In ALESS9.1, the CO(4--3) does not peak where the 870$\mu$m dust continuum does, instead showing a $\simeq$0.1$''$ spatial offset. This could point to an interaction or an optical depth effect.

\subsection{Molecular gas morphology}
\label{sec:mgm}

\begin{figure*}
\centering
\includegraphics[width=0.68\columnwidth]{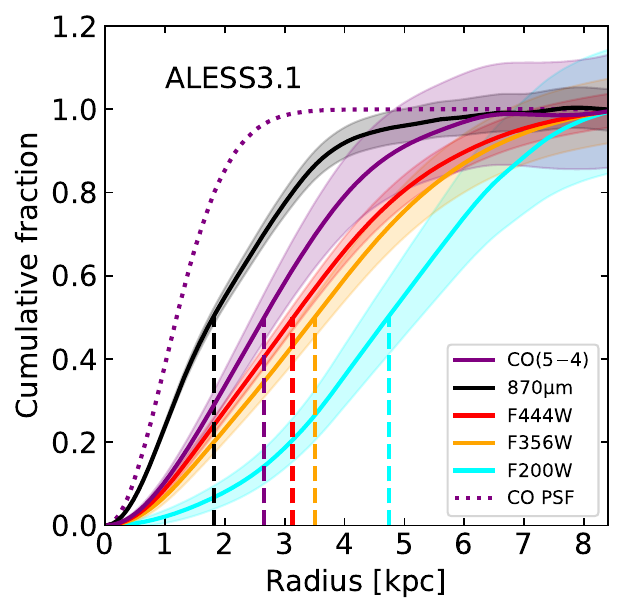}
\includegraphics[width=0.68\columnwidth]{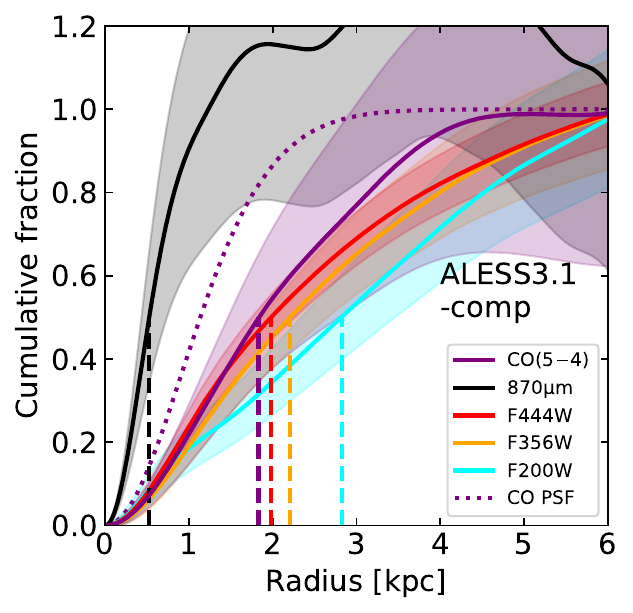}
\includegraphics[width=0.68\columnwidth]{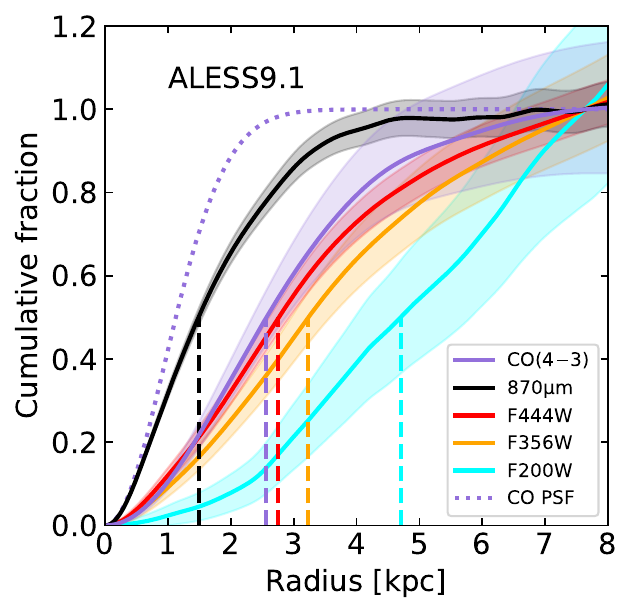}
\caption{Flux density profiles of CO line emission, dust continuum \protect\citep{hodge_alma_2019} and stellar continuum traced by NIRCam filters \protect\citep{hodge_aless-jwst_2025}.
Dashed lines correspond to effective radii, $R_{\rm{e}}$. Dotted lines correspond to the CO PSF.}
\label{fig:JWSTCOG}
\end{figure*}

\begin{table*}
    \centering
    \caption{Effective radii, $R_{\rm{e}}$ in kpc, obtained from curve-of-growth analysis of the CO emission lines, ALMA 870$\mu$m continuum map and NIRCam images. $R_{\rm{e,CO}}$ is for CO(5--4) for ALESS3.1 and ALESS3.1-comp and for CO(4--3) for ALESS9.1. The F444W and $870\,\rm{\mu m}$ values are from \protect\cite{hodge_aless-jwst_2025}.}
    \begin{tabular}{l|ccccc}
        \hline
        Source ID & $R_{\rm{e,CO}}$ [kpc] & $R_{\rm{e,870\mu m}}$ [kpc] & $R_{\rm{e,F444W}}$ [kpc] & $R_{\rm{e,F356W}}$ [kpc] & $R_{\rm{e,F200W}}$ [kpc] \\
        \hline
        ALESS3.1 & 
        $2.66^{+0.26}_{-0.21}$ & $1.82^{+0.08}_{-0.08}$ & $3.13^{+0.16}_{-0.13}$ & $3.50^{+0.26}_{-0.24}$ & 
        $4.74^{+0.53}_{-0.42}$ \\
        ALESS3.1-comp & 
        $1.83^{+0.96}_{-0.36}$ & $0.53^{+0.22}_{-0.09}$ & $1.97^{+0.29}_{-0.22}$ & $2.21^{+0.43}_{-0.33}$ & 
        $2.83^{+0.63}_{-0.49}$ \\
        ALESS9.1 & 
        $2.56^{+0.31}_{-0.22}$ & $1.50^{+0.05}_{-0.07}$ & $2.75^{+0.19}_{-0.14}$ & $3.23^{+0.34}_{-0.27}$ & 
        $4.70^{+1.11}_{-0.72}$ \\
        \hline
    \end{tabular}
    \label{tab:Re}
\end{table*}
We conduct a curve-of-growth analysis to determine the effective radii of ALESS3.1, ALESS3.1-comp and ALESS9.1 in each tracer. 

We use matching elliptical apertures to \cite{hodge_aless-jwst_2025}. The shape of these apertures and maximum sizes of the major and minor axes were determined by running \textsc{sextractor} \citep{bertin_sextractor_1996} on the NIRCam F444W image and the centroid as the peak of the 870$\mu$m emission from \cite{hodge_alma_2019}. We use \textsc{photutils} \citep{bradley_astropyphotutils_2022} to measure flux density as a function of distance along the major axis, with the total integrated flux density for each tracer taken as the flux density from the maximum aperture. The curves are not deconvolved from their point-spread functions (PSFs) but this has only a minor impact as the sources are well-resolved in all the tracers that we investigate. We show a curve corresponding to the CO PSF as an example as it has the lowest resolution. The effective radii, $R_{\rm{e}}$, found are given in Table~\ref{tab:Re}, with the NIRCam and ALMA 870$\mu$m the same as in \cite{hodge_aless-jwst_2025}. The flux density profiles are shown in Fig.~\ref{fig:JWSTCOG}.

We find that the CO-traced molecular gas is more extended than the $870 \mu m$ dust continuum for all three sources examined, with effective radii $\sim 1.5^{+0.2}_{-0.2}$, $3.5^{+2.9}_{-1.5}$ and $1.7^{+0.3}_{-0.2}$ $\times$ larger for ALESS3.1, 3.1-comp, and 9.1, respectively. This produces an average ratio of $R_{\rm{e,CO}}/R_{870\mu\rm{m}}=2.2^{+0.6}_{-0.5}$.
Previous high spatial resolution studies at high redshift have also shown that molecular gas is more extended than the 870$\mu$m dust emission in SMGs. For example, high-resolution ALMA studies found the dust continuum to be more compact than CO(3--2) emission \citep[e.g.,][] {chen_spatially_2017,calistro_rivera_resolving_2018,tadaki_spatial_2023}.  
Considering lower-$J$ CO transitions, \cite{rybakCO10ImagingReveals2025} use a stacking analysis of 19 SMGs with low-resolution CO(1--0) observations and find that their CO(1--0) reservoirs are $2-3\times$ more extended than the 870$\mu$m continuum on average. This is broadly consistent with our findings despite the fact that the CO(5--4) and CO(4--3) we observe are mid-$J$ CO transitions and may be expected to trace gas that is more compact, denser and warmer than CO(1--0). The difference we see between the extents of the molecular gas and 870$\mu$m dust continuum is potentially due to gradients in the dust temperature and optical depth within the sources and not a difference in the actual distribution of gas and dust \citep[][]{calistro_rivera_resolving_2018,boogaard_resolving_2026}.
Meanwhile, we find that the CO-traced molecular gas has a similar spatial extent to the rest-frame near-infrared emission detected in the F444W filter for all three galaxies ( $R_{\rm{e,CO}}/R_{\rm{e,F444W}}=0.9^{+0.2}_{-0.1}$) a trend also reported in \cite{tadaki_spatial_2023}.   

\begin{table}
    \centering
    \caption{S\'ersic modelling results for the high-resolution CO lines}
    \begin{tabular}{lccc}
        \hline
        Source ID & $n$ & $b/a$ & PA (deg) \\
        \hline
        ALESS3.1 & $0.73\pm0.08$ & $0.78\pm0.09$ & $144\pm16$ \\
        ALESS3.1-comp & $2.9\pm0.8$ & $0.37\pm0.08$ & $84\pm9$ \\
        ALESS9.1 & $0.74\pm0.08$  & $0.72\pm0.07$ & $61\pm9$ \\
        \hline
    \end{tabular}
    \label{tab:sersic}
\end{table}

We conduct further investigation of the gas morphology by fitting a S\'ersic model to the high-resolution zeroth moment CO maps, using \textsc{Astropy Sersic2D} and \textsc{Petrofit}  \citep{geda_petrofit_2022}. We show the resulting parameters in Table \ref{tab:sersic}. The S\'ersic models are convolved with the PSF of the beam (with the PSF profile shown in Fig.~\ref{fig:JWSTCOG}). The uncertainties on the resulting parameters of the S\'ersic model are first obtained using the covariance, setting the \textsc{Petrofit} fit\_model parameter calc\_uncertainties to True. We note the fitting does not account for spatially correlated noise which would potentially cause the uncertainties to be underestimated and so, following \cite{hodge_aless-jwst_2025}, we add 10\% fractional uncertainty in quadrature on all parameters.
We find that the effective radii from the S\'ersic modelling are consistent within 2$\sigma$ with the non-parametric $R_{\rm{e,CO}}$ values in Table \ref{tab:Re} so the latter are taken as the fiducial values given the lack of model dependence.
We also find that ALESS3.1 and ALESS9.1 have S\'ersic indexes of $n \sim $0.7. These values
indicate profiles that are closer to an exponential disc than a bulge, with indices similar to those derived for these sources from both the 870$\mu$m continuum emission and F444W light profiles using \textsc{galfit} by \citet{hodge_aless-jwst_2025}. The S\'ersic index for ALESS3.1-comp, meanwhile, is higher (as also seen in the F444W emission). Finally, we find values for the ellipticity ($b/a$) and position angle (PA) for all sources that are broadly consistent with the values found for the  870$\mu$m continuum emission and F444W emission in \cite{hodge_aless-jwst_2025}, indicating the three components are all tracing the same general structure. 

\subsection{Kinematic Modelling}
\label{sec:KinematicModelling}

For ALESS3.1 and ALESS9.1, the S/N and resolution of the CO data are sufficient to model the gas kinematics ($\rm{S/N}>10\,,\sim2\,\rm{kpc}$). For this purpose, we use \textsuperscript{\texttt{3D}}\texttt{BAROLO}  \citep{teodoro_3d_2015} to fit 3D tilted-ring models to the full emission-line data cubes whilst accounting for beam-smearing using forward modelling.
\textsuperscript{\texttt{3D}}\texttt{BAROLO} has been extensively tested and proved reliable at similar resolution to the CO data in this work \citep[e.g.][]{teodoro_3d_2015,mancerapina_robust_2020,rizzo_dynamical_2022,mancerapinaKinematicScalingRelations2026}.
We model the morphology using the \textsuperscript{\texttt{3D}}\texttt{BAROLO} sister code \texttt{CANNUBI}\footnote{https://www.filippofraternali.com/cannubi} \citep{roman-oliveira_regular_2023}, which estimates the inclination (\textit{i}) and position angle (PA) values for use with \textsuperscript{\texttt{3D}}\texttt{BAROLO}\footnote{We note that for ALESS3.1, the inclination failed to converge due to the irregular shape of the CO emission, and we thus use the inclination of the F444W emission \citep{hodge_aless-jwst_2025}}. The inclinations used are \textit{i}=$53\pm7^\circ$ (ALESS3.1) and \textit{i}=$46\pm17^\circ$ (ALESS9.1), while the PAs are PA=$160\pm29$ (ALESS3.1) and PA=$53\pm27^\circ$ (ALESS9.1). 
The \texttt{CANNUBI} values are consistent with both those derived from the S\'ersic fitting (Section~\ref{sec:mgm}) as well as the values measured from the JWST F444W emission \citep{hodge_aless-jwst_2025} within the uncertainties.

For the modelling, we select a $2\arcsec \times 2\arcsec$ subcube for each galaxy that spans $\pm$1.5\,$\times$\,FWHM of the CO emission line. The radial separation of the rings is $\sim$70 percent of the FWHM of the synthesized beam, allowing us to trace the kinematics with three largely independent beams. For this work, this means that the maximum number of independent rings possible is three, which has been shown to be reasonable to recover accurate values for rotation velocity and dispersion \citep[e.g.][]{mancerapina_robust_2020,rizzo_alma-alpaka_2023}. We choose to use the SEARCH source-finding parameter with a primary SNRCUT of 3${\sigma_{\rm{rms}}}$ above the mean and a GROWTHCUT of 2.5${\sigma_{\rm{rms}}}$ around the detected source. We then enlarge the mask spatially (+0.2--0.32$''$) and spectrally (+2 channels, ~40$\rm{km\,s^{-1}}$) until it encloses all the CO emission visible in the channel maps and PV diagrams. This approach allows us to consider low S/N emission without including significant background noise (see also \citealt{mancerapinaKinematicScalingRelations2026}). We minimise $\chi^2$ and use azimuthal normalisation for all fits. We fix the thickness of the galaxy to zero, as in \cite{roman-oliveira_regular_2023} and \cite{rowlandREBELS25DiscoveryDynamically2024a}. 
We explored leaving $V_{\rm{rad}}$ as a free parameter but we found no compelling evidence for radial motions. Therefore, $V_{\rm{rad}}$ is also set to zero so no radial motion is considered.
$V_{\rm{sys}}$ is estimated by \textsuperscript{\texttt{3D}}\texttt{BAROLO} as the central velocity in the global line profile and fixed to this value. We use the position of the peak of the second moment map as the kinematic centre. 
We leave the rotational velocity $V_{\rm{rot}}$ and velocity dispersion ${\sigma}$ as free parameters for \textsuperscript{\texttt{3D}}\texttt{BAROLO} to fit.

\begin{table*}
    \centering
    \caption{Kinematic parameters from \textsuperscript{\texttt{3D}}\texttt{BAROLO} modelling.}
    \begin{tabular}{lccccc}
        \hline
        Source ID & $V_{\rm{max}}\,\rm{[km \,s^{-1}]}$& $\overline{\sigma} \,\rm{[km\,s^{-1}]}$ & $V_{max}\,/\,\overline{\sigma}$ & $\tau_{\rm{orb}} [\rm{Myr}]$ & $M_{\rm{dyn,tot}} \,[10^{11}\,\rm{M_\odot}]$\\
        \hline
        ALESS3.1 & $473^{+50}_{-51}$ & $82^{+11}_{-11}$ & $6^{+1}_{-1}$ & $35^{+5}_{-5}$ & $2.9^{+0.6}_{-0.6}$\\
        ALESS9.1 & $427^{+62}_{-84}$ & $80^{+13}_{-16}$ & $5^{+1}_{-1}$ & 
        $37^{+9}_{-6}$ &
        $2.3^{+0.7}_{-0.8}$ \\
        \hline
    \end{tabular}
    \label{tab:param}
\end{table*}

\begin{figure*}
    \centering
    \includegraphics[scale=0.42]{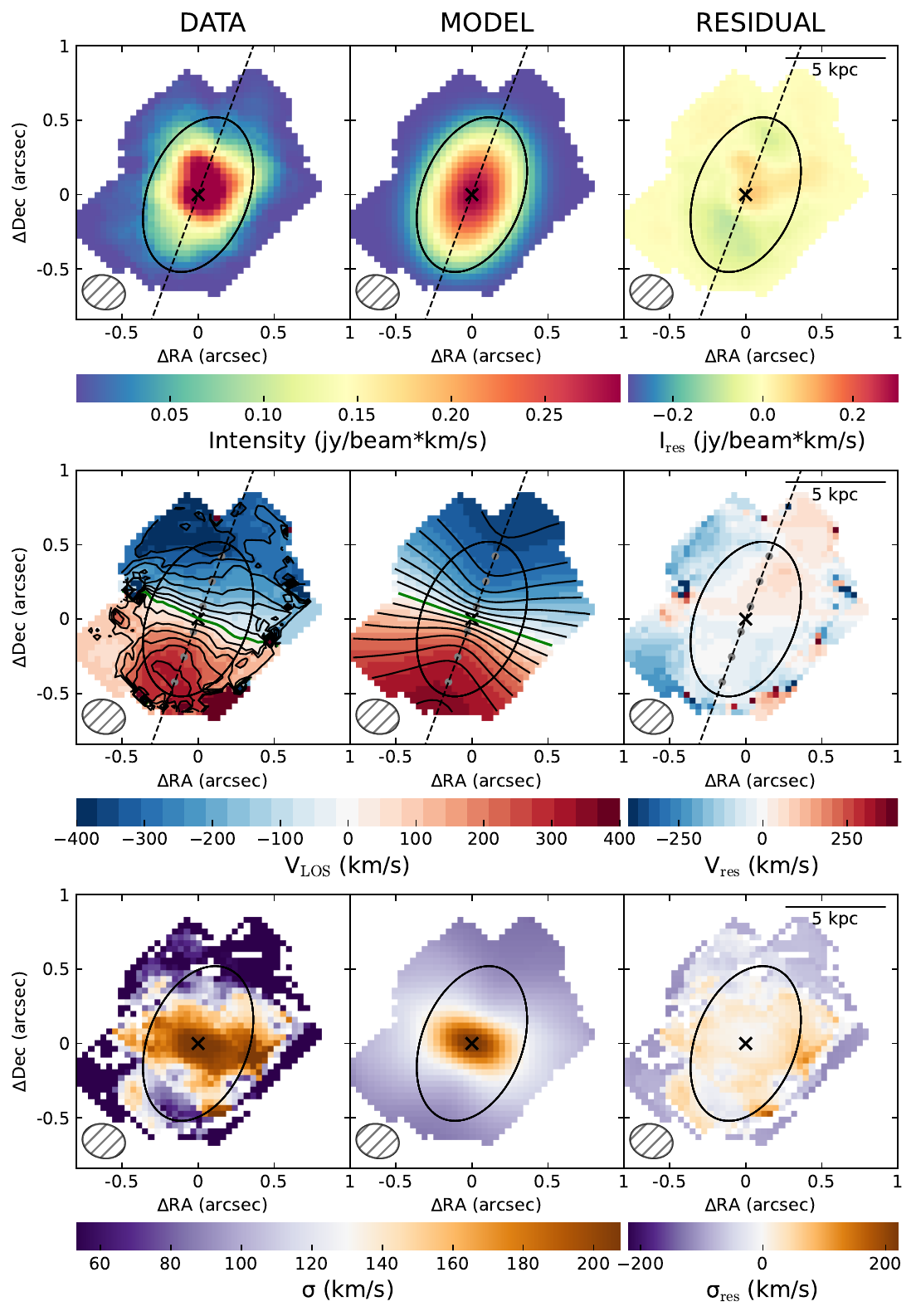}
    \includegraphics[scale=0.42]{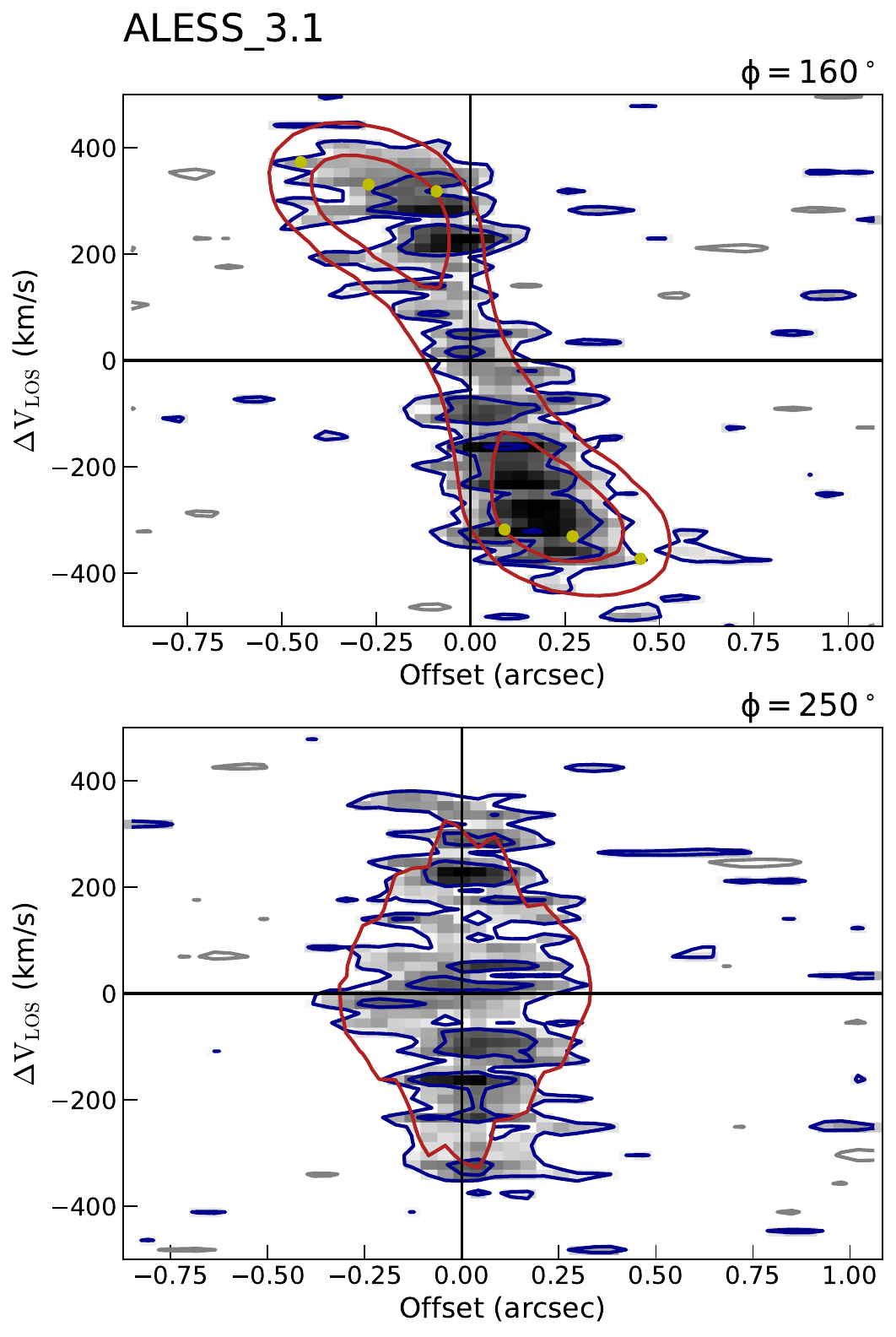}
    \includegraphics[scale=0.42]{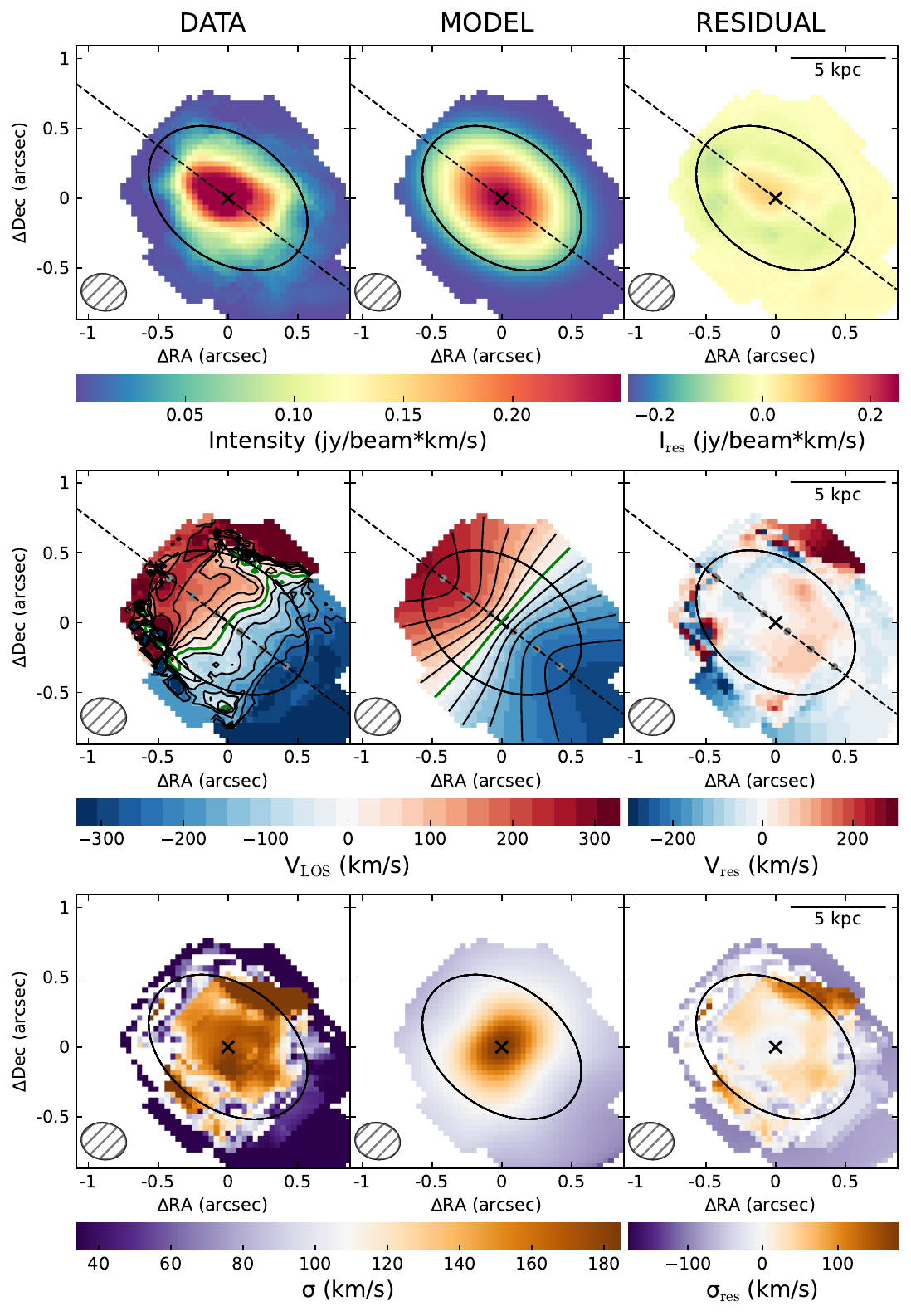}
    \includegraphics[scale=0.42]{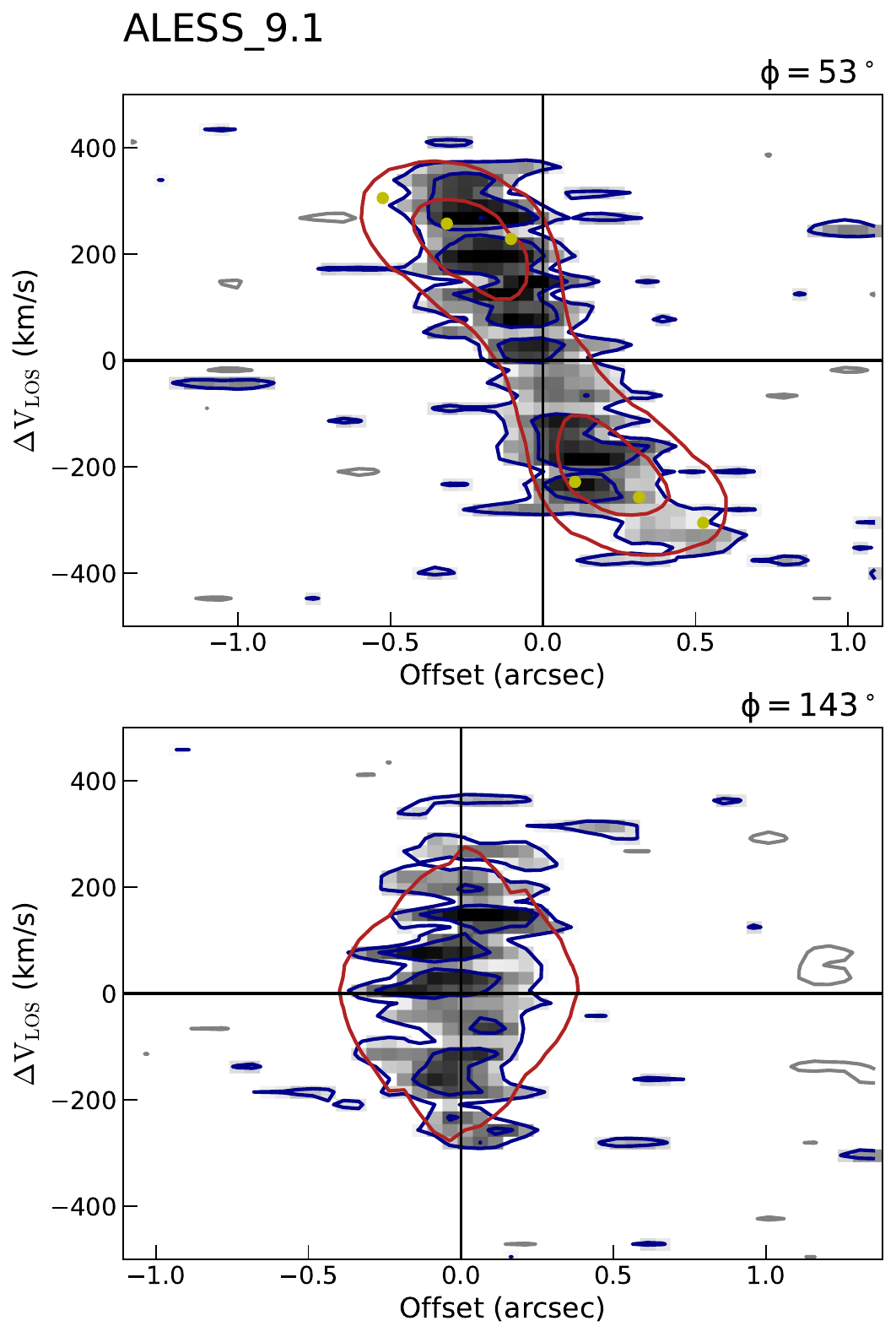}
    \caption{Kinematic modelling for ALESS3.1 (upper) and ALESS9.1 (lower): (Left) Moment 0, 1 and 2 maps with 50 km s$^{-1}$ velocity contours, model and beam ellipses (Right) position-velocity (PV) diagrams of the minor and major axes with minimum contours at 2$\sigma_{\rm{rms}}$, increasing in steps of 2$\sigma_{\rm{rms}}$. The best-fit kinematic model is shown with red contours and the data are shown with positive blue contours and negative grey contours. The yellow markers show $V_{\rm{rot}}$ of each ring projected along the line of sight. We find that the galaxies are well fit by rotating disc models at the current sensitivity and spatial resolution.
}
    \label{fig:3DB}
\end{figure*}

We take $V_{\rm{max}}$ as $V_{\rm{rot}}$ of the outermost ring and $\overline{\sigma}$ as the average ${\sigma}$ of the rings. The outputs from the modelling with \textsuperscript{\texttt{3D}}\texttt{BAROLO} is shown in Fig.~\ref{fig:3DB}
(with the $V_{\rm{max}}$ and $\overline{\sigma}$ parameters indicated in Table~\ref{tab:param}). We also calculate the orbital dynamical timescales at the effective radius, $\tau_{\rm{orb}}=2\pi R_{\rm{e,CO}}/V_{\rm{max}}$, showing the results in Table \ref{tab:param}.

In Fig.~\ref{fig:3DB}, we see a clear velocity gradient in the first moment maps for both galaxies, which can be indicative of rotation (or unresolved merging). For both galaxies, we also see the position velocity diagrams have the characteristic shape of a rotating disc.
This modelling therefore suggests that ALESS3.1 and ALESS9.1 are well-fit by rotating discs. The galaxies have high ratios of ordered-to-random motion, with $V_{\rm{max}}\,/\,\overline{\sigma}$ $\gtrsim$ 5. These ratios are somewhat conservative as we set BWEIGHT, the exponent of weight for blank pixels, to zero for the \textsuperscript{\texttt{3D}}\texttt{BAROLO} modelling which allows for larger $\sigma$  values than the default BWEIGHT=1 \citep{teodoro_3d_2015}. We note that excluding the velocity dispersion for the inner ring, which may include a non-negligible contribution from the rising part of the rotation curve, from the calculation of $\overline{\sigma}$ does not alter our values of $V_{max}\,/\,\overline{\sigma}$ beyond the quoted uncertainties.

We also explored whether the dusty substructures exhibit kinematics distinct from the overall rotating disc. Using local normalisation, which forces the model to reproduce the observed clumpiness more closely, we find that the inferred values of $V_{\rm rot}$ and $\overline{\sigma}$ remain consistent within the uncertainties with those derived from the azimuthally averaged model for both ALESS3.1 and ALESS9.1. While this suggests that the current data are broadly consistent with the dusty substructures participating in the overall disc rotation, we note that subtle local kinematic deviations may remain unresolved at the present sensitivity and angular resolution. 

Finally, we are able to estimate the total dynamical masses of ALESS3.1 and ALESS9.1. The dynamical mass, $M_{\rm{dyn,tot}}$, reflects the sum of all the mass components of the galaxy, primarily gas mass, stellar mass and dark matter. We first calculate the circular velocity, $V_{\rm{circ}}$, which includes an asymmetric drift correction, $V_{\rm{A}}$, to the rotational velocity of the outermost ring, $V_{\rm{max}}$, to account for pressure support, which is prominent in turbulent high-redshift galaxies \citep[][]{birkin_kaoss_2024,danhaiveDawnDiscsUnveiling2025,mancerapinaKinematicScalingRelations2026}. 

Following  \cite{iorio_little_2016}, we calculate
\begin{equation}
V_{A}^2 = -R\sigma_{v}^2\frac{\partial\ln{(\sigma_{v}^2}\Sigma_{\rm{int}})}{\partial R}
\end{equation}
\begin{equation}
V_{\rm{circ}}^2=V_{\rm{max}}^2+V_{\rm{A}}^2
\end{equation}
$R$ is the radii of the rings used for the \textsuperscript{\texttt{3D}}\texttt{BAROLO} modelling and $\sigma_{v}$ is the velocity dispersion at each ring. $\Sigma_{\rm{int}}$ is the intrinsic surface density and we find this through using the amplitude, effective radii and S\'ersic indicies of the 2D S\'ersic profiles of the galaxies (see Section \ref{sec:mgm}) to produce a 1D S\'ersic profile (using Sersic1D). We then make 1000 realisations of this profile to Monte Carlo sample and obtain $\Sigma_{\rm{int}}$ at each R. We also calculate $V_A$ for each ring also using a Monte Carlo approach, with 1000 realisations of $\sigma_{v}$ and $\Sigma_{\rm{int}}$ as input for Equation 2, taking $V_A$ as the median value calculated and the uncertainty on $V_A$ as 1.48 $\times$ the median absolute deviation. For simplicity, we use the $V_A$ of the outermost \textsuperscript{\texttt{3D}}\texttt{BAROLO} ring to calculate $V_{\rm{circ}}$. 

We then estimate the total dynamical mass following \citet{price_kinematics_2022}, and show the results in Table \ref{tab:param}:
\begin{equation}
M_{\rm{dyn,tot}}(R) = k_{\rm{tot}}\frac{RV^2_{\rm{circ}}(R)}{G}.
\end{equation}
We evaluate this expression at $R=R_{\rm{e,CO}}$ from Table \ref{tab:Re}. We adopt the pre-factor $k_{\rm{tot}}=1.8$ from \citet{price_kinematics_2022}, which relates the total system mass to the circular velocity for a thin disc model and is appropriate for the S\'ersic indices and intrinsic axis ratios of these sources.

The resulting dynamical masses are similar in scale to the stellar masses derived from the SED modelling, as shown in Table \ref{tab:targetinfo}. This similarity between the dynamical and stellar mass estimates leaves little room for a substantial additional gas component, implying a $2\sigma$ limit of $\alpha_{\rm CO}\lesssim 1.2$. A high conversion factor such as $\alpha_{\rm{CO}}=4.0\pm0.1\,\rm{M_\odot}\,(K\,km\,s^{-1}\,pc^2)^{-1}$, as suggested by \citet{dunne_dust_2022}, would therefore overpredict the gas mass allowed by the dynamical constraints.

These results support our adoption of a low conversion factor, $\alpha_{\rm{CO}}=1\,\rm{M_\odot}(K\,km\,s^{-1}\,pc^2)^{-1}$, for the calculation of the total gas mass in Section \ref{sec:SpectraAndImaging} and in the resolved annular regions in Section \ref{sec:SFL}. This choice is also consistent with previous dynamical studies \citep[e.g.][]{danielson_properties_2011,hodge_evidence_2012,bothwell_survey_2013, chen_spatially_2017, calistro_rivera_resolving_2018,xue_flat_2018,riechers_coldz_2020,amvrosiadis_kinematics_2025}, as well as independent constraints based on the solar or super-solar metallicities of SMGs \citep{frias_castillo_comparative_2025}. Similar conclusions have also been drawn for some SMGs in the KAOSS \citep{birkin_kaoss_2024} and FOSSILS \citep{ikeda_formation_2025} samples. These studies all support the interpretation that these galaxies are baryon dominated on $\lesssim$ 10 kpc scales.

The above estimates of dynamical mass carry uncertainties from the assumptions of a flat rotation curve and a single thin disc component consisting of stars and gas. We therefore also constrain $\alpha_{\rm{CO}}$ using an alternative approach that does not assume a specific rotation-curve shape and instead considers separate, thick stellar and gas disc components. This approach, described in Appendix \ref{sec:Mdyn}, independently indicates that $\alpha_{\rm{CO}}\lesssim1$ in our galaxies.

\subsection{Resolved star formation scaling relations}
\label{sec:SFL}

The resolved molecular gas data, combined with the high-resolution submillimetre continuum maps, provide an opportunity to investigate star formation scaling relations at high redshift on kiloparsec scales. To investigate these relations on resolved scales, we split the galaxies into four concentric rings surrounding a circular inner region with a diameter of 2 kpc, which is approximately the FWHM of the CO beam. The diameter of the outer ring of each region increases incrementally by 2 kpc. 

\begin{figure*}
    \centering
    \includegraphics[width=\textwidth]{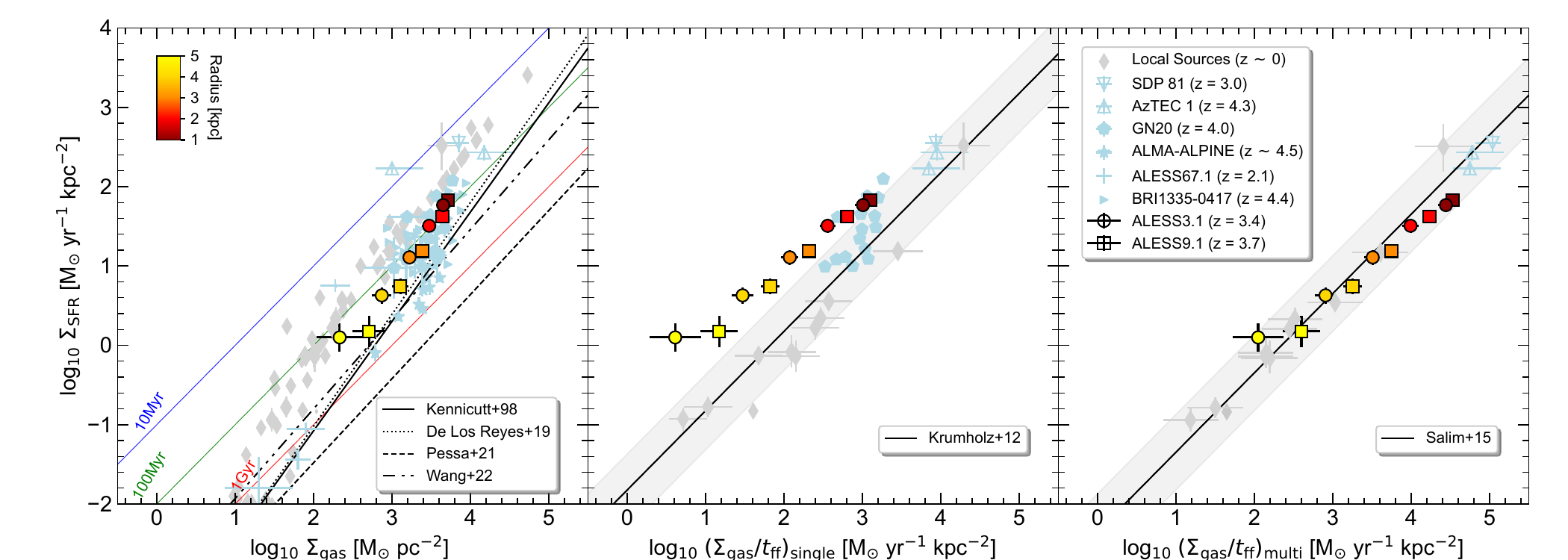}
    \caption{Star formation relation comparison showing annular regions for ALESS3.1 (circle) and ALESS9.1 (square) along with both resolved and unresolved literature values from SFGs and SMGs. Also plotted are the \protect\cite{kennicutt_global_1998}, \protect\cite{de_los_reyes_revisiting_2019}, \protect\cite{pessa_star_2021}, \protect\cite{wang_3_2022} relations (general Kennicutt-Schmidt relation, left panel), \protect\cite{krumholz_universal_2012} relation (incorporating free-fall timescale, centre panel) and \protect\cite{salim_universal_2015} relation (incorporating turbulence, right panel) with agreement between models and observations increasing from left to right. Local star forming regions are in grey (\protect\citealt{kennicutt_global_1998}, \protect\citealt{heiderman_star_2010}, \protect\citealt{lada_star_2010}, \protect\citealt{wu_properties_2010}, \protect\citealt{gutermuth_correlation_2011}, \protect\citealt{jameson_relationship_2016}, \protect\citealt{federrath_link_2016}) and high-redshift star forming regions are in blue (SDP81,  \protect\citealt{sharda_testing_2018}; AZTEC1,  \protect\citealt{sharda_testing_2019}; GN20,  \protect\citealt{hodge_kiloparsec-scale_2015}; ALMA-ALPINE, \protect\citealt{bethermin_alma-alpine_2023}; ALESS67.1, \protect\citealt{chen_spatially_2017} and BRI1335-0417, \protect\citealt{tsukui_spatially_2023}). For ALESS3.1 and ALESS9.1, the more inner radial regions are red whilst the more outer regions are yellow. The shaded regions show deviations by a factor of three from $\epsilon_{\rm{ff}}=0.015$ \protect\citep{krumholz_erratum_2013}.}
    \label{fig:KS}
\end{figure*}

The first collection of star formation relations we investigate are empirical relations between the gas surface density and the SFR surface density (e.g., the Kennicutt-Schmidt relation). For the gas surface density measurements 
($\Sigma_{\rm gas}$),
we use the high-resolution CO data, which for ALESS3.1 is CO(5--4) and for ALESS9.1 is CO(4--3). We first use the \textsc{CASA} \texttt{imstat} task to sum the CO line intensity $I_{\mathrm{CO,J}}$ in Jy km s$^{-1}$ within the relevant annular region on the zeroth moment map. We convert these values to gas masses assuming fixed values of the excitation ratio and $\alpha_{\rm CO}$, as described in Section~\ref{sec:SpectraAndImaging}. 
We finally calculate gas surface density 
$\Sigma_{\rm gas}$
in $\rm{M_\odot\,pc^{-2}}$ by dividing by the total area of the region in $\rm{pc}^2$. The error on 
$\Sigma_{\rm gas}$
is calculated from the errors on $I_{\mathrm{CO,J}}$ and the excitation ratio used ($r_{41}$ or $r_{51}$). 

We note that CO(5--4) has been suggested to be empirically related to the dense gas component \citep{daddi_co_2015}. We therefore also utilise an alternative method for obtaining the gas masses from the dust masses for each region calculated using modified black body modelling of the two high-resolution continuum detections, the $870\rm{\mu m}$ and the continuum underlying the CO, and again using the SMG gas-to-dust-ratio \citep[$\delta_{\rm{gdr}}/\alpha_{\rm{CO}}=63\pm7$,][]{birkin_almanoema_2021}. For this modelling, we use an optically thin dust approximation and fix the dust emissivity index to $\beta=1.9$ as derived for ALESS3.1 and ALESS9.1 in \cite{da_cunha_measurements_2021}. For ALESS9.1, the data are insufficient to constrain the dust temperature, but for ALESS3.1 this method provides gas mass surface densities that are consistent with those derived from the CO above (see Appendix \ref{sec:MBB}).

For the SFR surface density measurements ($\Sigma_{\rm SFR}$), we use high-resolution 
870$\mu$m dust continuum data from \cite{hodge_alma_2019}. We produce a global ratio by dividing the total SFR from integrated SED fitting (\citealt{li_aless--jwst_2026}; Table~\ref{tab:targetinfo}) by the total 870$\mu$m flux density $S_{870}$ (\citealt{hodge_aless-jwst_2025}, Table~\ref{tab:targetinfo}). We convolve the 870$\mu$m dust map to the beam of the CO map using \texttt{imsmooth}. We calculate $I_{\mathrm{dust}}$ by summing the intensity of each pixel within the region on the dust map using the \textsc{CASA} \texttt{imstat} task. We then find the SFR for the region by multiplying $I_{\mathrm{dust}}$ by the global ratio. We calculate the star-formation rate surface density, $\Sigma_{\rm SFR}$, by dividing by the total area of the region in $\rm{kpc}^2$. The error on $\Sigma_{\rm SFR}$ is calculated in quadrature from the error on $I_{\mathrm{dust}}$ and the error on the global ratio. We also test an alternative method for deriving $\Sigma_{\rm SFR}$, using the modified black body modelling as described above, obtaining the SFR from dust luminosity using the conversion factor from \cite{murphy_calibrating_2011}, and the results are again consistent (see Appendix \ref{sec:MBB}). 

We present our $\Sigma_{\rm SFR}$ versus $\Sigma_{\rm{gas}}$ measurements in Fig.~\ref{fig:KS} (left panel). For comparison, we include a selection of resolved and unresolved observations across a range of redshifts \citep{kennicutt_global_1998,heiderman_star_2010,lada_star_2010,wu_properties_2010,gutermuth_correlation_2011,hodge_kiloparsec-scale_2015,jameson_relationship_2016,federrath_link_2016,chen_spatially_2017,sharda_testing_2018,sharda_testing_2019,bethermin_alma-alpine_2023,tsukui_spatially_2023}. We find that ALESS3.1 and ALESS9.1 generally lie in the same parameter space also occupied by local and high-redshift sources from both resolved and unresolved studies.

When fitted using orthogonal distance regression for a function of the form $\Sigma_{\rm{SFR}}\sim(\Sigma_{\mathrm{gas}})^N$, for ALESS3.1 we find $N=1.38\pm0.08$, and for ALESS9.1 we find $N=1.70\pm0.09$. The regions in ALESS9.1 therefore show a marginally steeper relation than the standard $N=1.4$ Kennicutt-Schmidt relation \citep[][corrected for a Chabrier IMF as in \citealt{sharda_testing_2018}]{kennicutt_global_1998}, whilst the relation for ALESS3.1 is consistent within the uncertainties. 
We find gas depletion times for these resolved regions within both galaxies to be $\sim$100\,Myr (ranging from 90 -- 400 Myr), which is consistent with other submillimetre-bright galaxies \citep[e.g.,][]{birkin_almanoema_2021, frias_castillo_vla_2023}.
We find that the centres of the galaxies have shorter depletion times than the outskirts, with the caveat that we assume constant dust temperature, dust opacity, $\alpha_{\rm CO}$ and $r_{\rm J,J-1}$ as a function of radius (see Sec \ref{KSdiscuss}).

Star formation is influenced not only by the quantity of gas but also by the time available to convert this gas into stars.
We therefore also compare to the star-formation law proposed by \cite{krumholz_universal_2012}, which is a notable example of a modified version of the Kennicutt-Schmidt law that incorporates the free-fall timescale of the gas:
\begin{equation}
    \Sigma_{\rm{SFR}}=f_{\rm{H_2}}\epsilon_{\rm ff}\frac{\Sigma_{\rm{gas}}}{t_{\rm ff}}
\end{equation}
where $\epsilon_{\rm{ff}}$ denotes the star formation efficiency per free-fall time. We assume that the fraction of star-forming gas, $f_{\rm{H_2}}=1$ for simplicity.

We compute free-fall time $t_{\rm ff}$ of each region in Myr using
\begin{equation}
    t_{\rm ff}=\sqrt{ \frac{3\pi}{32G\rho} } \,.
\end{equation}
In our calculation of the density $\rho$ of each concentric annular region, we use the spherical volume of the region: $\rho=\frac{M_{\mathrm{gas}}}{\frac{4}{3}\pi(r_2^3-r_1^3)}$ where $r_2$ denotes the outer radius of the ring and $r_1$ the inner. For the circular inner region, we use $\rho=\frac{M_{\mathrm{gas}}}{\frac{4}{3}\pi(r^3)}$ where $r$ is the 2 kpc radius.
We are limited by the resolution of the observations to using this spherical approximation but note that this concerns resolved regions within the disc rather than the entire disc itself. 
We find that values of $t_{\rm ff}$ vary across the galaxy from $\sim$ 4Myr in the centre to approximately 10$\times$ larger values in the outskirts. 
We show the results incorporating these free-fall times in Fig.~\ref{fig:KS} (centre panel). 
We see that our high-redshift data points are offset from the local relation. Both ALESS3.1 and ALESS9.1 are also found to be less steep than the unity relation from \cite{krumholz_universal_2012} (slopes of $0.73\pm0.03$ and $0.86\pm0.02$). The fraction of gas converted into stars per free-fall time ($\epsilon_{\rm ff}$) is significantly different from the $\epsilon_{\rm ff}$$=0.015$ fit by \cite{krumholz_erratum_2013}, with $0.10\pm0.02$ and $0.068\pm0.005$ for ALESS3.1 and ALESS9.1, respectively. We find that the outskirts have higher $\epsilon_{\rm ff}$ than the inner regions. 

Finally, we consider gravo-turbulent models, which emphasise the critical role of turbulence in regulating star formation processes 
\citep[e.g.][]{padoan_star_2011,federrath_star_2012,burkhart_star_2018,mocz_markov_2019,meidtReconcilingExtragalacticStar2025a}. Central to these models is the concept that turbulence in the ISM generates a distribution of gas densities, resulting in variations in both the rates and timescales of star formation across different densities \citep{federrath_comparing_2010}. Supported by robust kinematic data, we are now positioned to test these models in galaxies at high redshift (see also \citealt{sharda_testing_2018, sharda_testing_2019}). In this work, we adopt a turbulence-regulated star formation model from \cite{salim_universal_2015} that takes the gas density distribution into account when calculating the free-fall time of a collapsing molecular cloud:
\begin{equation}
    \Sigma_{\rm{SFR}}=\epsilon^{\prime}_{\rm ff}\frac{\Sigma_{\rm{gas}}}{t_{\rm ff}}\left(1+b^2 \mathcal{M}^2\frac{\beta_{\rm{mag}}}{\beta_{\rm{mag}}+1}\right)^{3/8}
\end{equation}
where, analogous to $\epsilon_{\rm{ff}}$ in the single free-fall model above, $\epsilon^{\prime}_{\rm{ff}}$ is the multi-freefall efficiency.

To account for turbulence, we first calculate the turbulent Mach number $\mathcal{M}$ as
\begin{equation}
\mathcal{M}=\frac{\overline{\sigma}}{c_s}\,.
\end{equation}
We use the $\overline{\sigma}$ value in $\rm{km\,s^{-1}}$ from the kinematic modelling (Table~\ref{tab:param}). We take the speed of sound, $c_{\rm{s}}$, to be $0.4\pm0.2\,\rm{km\,s^{-1}}$ as in \cite{sharda_testing_2018}, where the uncertainty arises from our assumption that the (cold) gas temperature is somewhere between $12 - 100\,\rm{K}$. For ALESS3.1, we calculate $\mathcal{M}=205\pm106$ and for ALESS9.1, $\mathcal{M}=200\pm105$. We use the Mach number to calculate the gas surface density per multi-free fall time in $\rm{M}_\odot$ yr$^{-1}$kpc$^{-2}$ following \cite{salim_universal_2015} as in \cite{sharda_testing_2019} by multiplying the gas density per single free-fall time by the multi-free fall time correction 
$
(1+b^2\mathcal{M}^2\frac{\beta_{\rm{mag}}}{\beta_{\rm{mag}}+1})^{3/8}
$
where $b=0.4$ for a mixed turbulent driving mode and $\frac{\beta_{\rm{mag}}}{\beta_{\rm{mag}}+1} \to 1$ neglecting the effects of magnetic fields. We show this in the rightmost panel of Fig.~\ref{fig:KS}. We find that the fraction of gas converted into stars per time for the multi free-fall model is $\epsilon^{\prime}_{\rm{multi-ff}}=(3.8\pm0.7)\times10^{-3}$ for ALESS3.1 and $\epsilon^{\prime}_{\rm{multi-ff}}=(2.5\pm0.2)\times10^{-3}$ for ALESS9.1, similar to $\epsilon^{\prime}_{\rm{ff}} = 4.5\times10^{-3}$ fit by \citet{salim_universal_2015}.

We see that with increasing model complexity, the agreement between the model and observations improves, even across several orders of magnitude on the abscissa. Our results suggest that for resolved kiloparsec-scale galaxies at high redshift, incorporating the effects of turbulence is critical. We discuss this further in Section~\ref{KSdiscuss}.

The excitation ratios used to convert from the mid-$J$ CO lines observed to CO(1--0) factor into where the data lies on Fig.\ref{fig:KS}. In this work, we use $r_{41}$=0.34$\pm$0.04 and $r_{51}$=0.36$\pm$0.07 from \cite{birkin_almanoema_2021}, which are derived from a large sample of dusty star-forming galaxies including sources from ALESS.  If other excitation values from the literature were used, for example those of the Cosmic Eyelash from \cite{danielson_properties_2011}  ($r_{41}$=0.50$\pm$0.04 and $r_{51}$=0.35$\pm$0.02), those from the \cite{bothwell_survey_2013} SMGs ($r_{41}$=0.41$\pm$0.07 and $r_{51}$=0.32$\pm$0.05), $r_{41}$=0.63$\pm$0.44 from the \cite{frias_castillo_vla_2023} SMGs or those from the Vz-GAL \cite{prajapatiGALProbingCold2026} SMGs ($r_{41}$=0.49$\pm$0.15 and $r_{51}$=0.47$\pm$0.13), our basic findings remain unchanged. This is because the alternative values are either consistent with the values for this work, or higher, which would have the effect of moving the galaxies further from both the standard Kennicutt-Schmidt relation and the \cite{krumholz_universal_2012} star formation law, but even closer to the \citet{salim_universal_2015} multi-free fall model that accounts for turbulence.

\section{Discussion}
\subsection{What is triggering dusty star formation in these sources?}

The CO data presented here reveal that ALESS 3.1 lies at the same redshift as its nearby companion galaxy ALESS3.1-comp. Given this, the high far-infrared luminosity of ALESS3.1 and that there is evidence for potential tidal features in the 870$\mu$m and NIRCam imaging, these galaxies are likely interacting or undergoing early-stage mergers. This is consistent with their visual classifications \citep{hodge_aless-jwst_2025}. ALESS9.1, on the other hand, has no spectroscopically confirmed companion, but it does show a potential tidal feature in the 870$\mu$m and NIRCam imaging. It is therefore possibly in an advanced interaction or merger with one of the nearby galaxies at a similar photometric redshift \citep{alberts_high_2024}, as also discussed in \cite{hodge_aless-jwst_2025}. 

Despite the above evidence for interaction/merger activity, we nevertheless find in Section \ref{sec:KinematicModelling} that the cool gas kinematics of ALESS3.1 and ALESS9.1 are consistent with disc rotation without strong non-circular motions. Fig.~\ref{fig:ALMACOG} shows that ALESS3.1 and ALESS9.1 exhibit double-horn profiles with moderately asymmetric amplitude; we note that asymmetric profiles are commonly observed in most regularly-rotating nearby discs and satellites, and do not impact their dynamics significantly \citep[e.g.,][]{watts_global_2020,yu_statistical_2022}. In addition, while the high-resolution 870$\mu$m continuum in both galaxies shows distinct substructure, we see no evidence that these dusty substructures have kinematics that deviate from the strong bulk rotation. This may imply that the 870$\mu$m-detected substructures are coherent (lending support for dusty bars/rings/spiral arms) rather than representing, e.g., distinct merging components.
We note that based on the short dynamical timescales we estimate for these galaxies, there could have been a merger or interaction that triggered the current burst of dusty star formation as recently as $\sim$ 40\,Myr ago (see Table \ref{tab:param}) and the kinematics might still have had sufficient time to resettle into discs.

Interestingly, despite the evidence for recent interactions, we also find that the galaxies are dynamically cold with $V_{\rm{max}}\,/\,\overline{\sigma}$ $\gtrsim$ 5. This is consistent with values for $z=0$ gas-rich galaxies with moderate masses \citep{mancera_pina_galaxy-halo_2025}. 
Notably, ALESS3.1 and ALESS9.1 possess high stellar masses ($\log (M_{\star}/\rm{M_{\odot}}) > 11$).
\cite{kohandel_dynamically_2024} suggest from their theoretical models that the most-massive galaxies at $z\gtrsim5$ will display greater rotational support compared to less-massive galaxies. It is plausible that these galaxies, potentially accreting mass during ongoing interactions, will dynamically settle further as their mass increases. Such evolution might contribute to a reduction in their velocity dispersions and further reinforce their dynamically cold nature. 

Overall, given the identification of nearby companions and morphological signatures of interactions in the NIRCam and 870$\mu$m imaging, together with robust fits to rotating disc models and high ratios of ordered-to-random motion, these sources illustrate that dynamically cold discs traced by cool gas can, in some cases, coexist with signs of recent interactions. In other words, it is important to emphasize that the presence of a disc does not preclude a dusty star forming galaxy from having a merger or interaction origin, especially for systems at high redshift with substantial gas fractions. Such observations reinforce the notion that merger-driven assembly and dynamically cold disc formation are not incompatible in the early Universe.

\subsection{Testing star formation relations at $z\sim3$}\label{KSdiscuss}

In Section~\ref{sec:SFL}, we examined spatially resolved star formation relations on kiloparsec scales for ALESS3.1 and ALESS9.1 and found that the data are offset from the star formation relation from \citet{krumholz_universal_2012} and more consistent with the \citet{salim_universal_2015} model, which accounts for turbulence. Why might this be? In the \citet{krumholz_universal_2012} star formation relation, accounting for the average free-fall time in each region means that the differing three-dimensional sizes and internal clumping are considered. However, gravo-turbulent models take this a step further, incorporating the critical importance of turbulence in defining how stars form by incorporating a multi-freefall correction factor. In effect, these laws consider that gas with greater density has a shorter free-fall time, and thus forms stars at a higher rate. 
One advantage of multi-free fall models is that the large dynamic range they allow can be better suited at $z>2$ in gas-rich starburst environments \citep{meidtReconcilingExtragalacticStar2025a}. Indeed, we find that ALESS3.1 and ALESS9.1 lie close to the \citet{salim_universal_2015} relation. 

Further, it is interesting to note that the two formalisms lead to qualitatively different interpretations. Our derived value of the single-freefall efficiency is $\approx 10\times$ larger than the canonical value of $\sim$1 percent, whereas we find our multi-freefall efficiency to be the similar to the canonical value of $\sim$0.5 percent. The single-freefall model therefore suggests that star formation proceeds more efficiently in ALESS 3.1 and 9.1, whereas the multi-freefall model suggests that star formation in these galaxies is similar to that in the local Universe once the turbulent density structure of the star-forming gas is accounted for. In other words, we find that turbulence reduces the net star formation efficiency relative to the standard case. However, this should not be interpreted as a universal suppressive effect of turbulence: depending on the properties of the density PDF, turbulence can both inhibit and enhance star formation \citep[e.g.][]{federrath_link_2016,sharda_first_2021}.

We note that our resolution is not high enough to resolve individual star-forming clumps (100--200 pc), which may impact the interpretation of velocity dispersion. In particular, it is possible that the Mach numbers we measure on kiloparsec scales are greater than they would be on cloud scales as the velocity dispersions are effectively measured at the molecular cloud complex scale. Specifically, the Mach numbers measured for Milky Way molecular clouds are only of order $\sim$25 (compared to $\sim$200 for our sources). However, more vigorous star-formation causes more violent gas turbulence, so the Mach numbers would be expected to be greater for more highly star-forming galaxies like SMGs. Indeed, our measured Mach numbers are consistent with those of local ULIRGs ($\sim$200; e.g., \citealt{genzel_study_2010}).

We also note that the star formation relations tested in this work that account for free-fall are bottom-up models, meaning that they are built from smaller cloud scales to explain star formation rates, although they still yield valuable insights when applied to larger scales. 
There are also examples of top-down models, which instead go from larger (galaxy) scales to smaller (cloud) scales
\citep[e.g.][]{faucher-giguere_feedback-regulated_2013,ostriker_pressure-regulated_2022}. Another example of top-down models are the volumetric star formation relations, such as \cite{bacchini_volumetric_2019}, which require an estimation of the scale height of the disc, which is in turn set by the turbulence of the ISM. Observationally estimating these factors is outside the scope of this work and unfeasible at the resolution of the observations. Nevertheless, both the bottom-up and top-down models point to turbulence playing an important role in modulating the star formation rate in galaxies. 

In addition to the above considerations, there are several more general assumptions that underpin these results. One of these is the excitation ratios, for which, as discussed in Section \ref{sec:SFL}, our conclusions would not change if we used alternative values. An additional assumption is the adoption of a CO-to-H$_2$ conversion factor. 
The value of $\alpha_{\rm{CO}}=1\,\rm{M_\odot}\, (K\,km\,s^{-1}\,pc^2)^{-1}$ was chosen to be consistent with the literature on SMGs \citep[e.g.][]{calistro_rivera_resolving_2018,birkin_almanoema_2021,frias_castillo_kiloparsec-scale_2022,taylor_properties_2024,amvrosiadis_kinematics_2025}.
If we had instead used a much larger value of $\alpha_{\rm{CO}}$, for example $\alpha_{\rm{CO}}=4.0\pm0.1\,\rm{M_\odot}\, (K\,km\,s^{-1}\,pc^2)^{-1}$  from \cite{dunne_dust_2022}, our data would then become consistent with the \cite{krumholz_universal_2012} star formation relation without the need to incorporate additional model complexity. 
However, the \citet{dunne_dust_2022} value is in tension with previous estimates based on resolved dynamics of SMGs \citep{hodge_evidence_2012, calistro_rivera_resolving_2018, amvrosiadis_kinematics_2025}. This, combined with our own limits based on the galaxies' dynamical masses (as discussed in section \ref{sec:KinematicModelling}), leads us to prefer $\alpha_{\rm{CO}}\sim1\,\rm{M_\odot}\, (K\,km\,s^{-1}\,pc^2)^{-1}$ as the most suitable conversion factor for this study, with the acknowledgement that any choice of $\alpha_{\rm{CO}}$ has significant impact on such tests of the star formation relations.  

Finally, our analysis necessarily assumes constant excitation ratios, $\alpha_{\rm{CO}}$ conversion factors and dust temperatures throughout the (resolved) galaxies. It is likely that these factors vary within the galaxies, which would impact the resolved star formation relations derived. For example, if we would take the dust temperature to be hotter in the centre of the galaxies than in the outskirts \citep[e.g.,][c.f. \citealt{rybak_full_2020}]{calistro_rivera_resolving_2018, tsukui_spatially_2023, boogaard_resolving_2026}, the relations would steepen due to a greater range of $\Sigma_{\rm{SFR}}$. If we would assume that the excitation ratio changes within the galaxy, as it may if the CO(1--0)--traced reservoir is more extended than mid-$J$ CO reservoirs \citep[e.g.,][]{ivison_tracing_2011}, this would steepen the relation due to a smaller range of $\Sigma_{\rm{gas}}$. Lastly, if we would assume a lower $\alpha_{\rm{CO}}$ in the centres \citep[e.g.,][]{sandstrom_co--h2_2013,teng_physical_2023}, this may also steepen the relation (again due to a smaller range of $\Sigma_{\rm{gas}}$). These effects may therefore affect the conclusions relating to the differences in depletion times and $\epsilon_{\rm ff}$ between the centres and outskirts of our sources. 
While testing these assumptions is beyond the scope of this work, they can begin to be addressed with resolved, multi-band and/or multi-$J$ data (e.g. the future multi-band 1\,kpc-scale dust continuum ALESS-JWST study 2025.1.00761.S [PI:Westoby]).

With the assumptions we have taken, the data presented here therefore suggest that accounting for turbulence is important, even in galaxies that appear dynamically cold.  
Cosmological hydrodynamic simulations \citep[e.g.,][]{kraljic_emergence_2024} emphasise how both environmental (mergers) and internal (gas fractions) conditions leave an imprint on star formation relations.
Stellar feedback, galactic dynamics and interaction-triggered stirring are all potentially driving the turbulence inside these galaxies (e.g., \citealt{krumholz_unified_2018}, \citealt{ginzburg_evolution_2022}).
Future studies of larger samples of galaxies with lower-$J$ CO transitions (e.g., the ALMA CONDOR Large Program 2024.1.00100.L [PI: Rizzo]) will help shed further light on the role of turbulence at high redshift. 

\section{Conclusions}

We present CO(5--4) and CO(4--3) ALMA imaging for three submillimetre-bright, dusty star-forming galaxies from the ALESS sample: ALESS3.1, ALESS3.1-comp and ALESS9.1. These data complement existing high-resolution ALMA 870µm continuum imaging \citep[][]{hodge_kiloparsec-scale_2016, hodge_alma_2019} and JWST NIRCam and MIRI imaging from the ALESS-JWST program \citep[][]{hodge_aless-jwst_2025,li_aless--jwst_2026}. This resolved multi-tracer study sheds new light on the properties of these galaxies and helps to bridge the gap between models and observations. We map their molecular gas morphology on kiloparsec scales and, for the two sources with the highest S/N CO observations (ALESS3.1 and ALESS9.1),  we conduct kinematic modelling and use them to investigate resolved star formation relations at high-redshift. Our findings are:
\begin{list}{\textbullet}{\setlength{\leftmargin}{1.5em}}
\item 
We spectroscopically confirm the redshift of ALESS3.1-comp 
and find it to be close to the redshift of ALESS3.1, indicating this system is possibly caught in an ongoing interaction.
\\ 
\item 
We find that the CO-traced molecular gas broadly follows the morphology of the 870$\mu$m dust continuum. Notably, dusty substructures, such as the northwestern arm/tidal feature in ALESS3.1, that were previously identified in the dust continuum and NIRCam imaging are also seen in CO.\\
\item Our resolved curve-of-growth analysis reveals that the molecular gas reservoirs are 1.5 to 3.5 times more extended than the dust continuum ($R_{\rm{e,CO}}/R_{870\mu\rm{m}}=2.2^{+0.6}_{-0.5}$), exhibiting a spatial extent comparable to the rest-frame near-infrared stellar emission seen in the NIRCam F444W imaging ( $R_{\rm{e,CO}}/R_{\rm{e,F444W}}=0.9^{+0.2}_{-0.1}$). The S\'ersic indices for ALESS3.1 and ALESS9.1 ($n \sim0.7$) indicate gas morphologies closer to an exponential disc rather than a bulge.\\
\item 
Our kinematic modelling with \textsuperscript{\texttt{3D}}\texttt{BAROLO} demonstrates that ALESS3.1 and ALESS9.1 are dynamically cold rotating discs ($V_{\rm{max}}\,/\,\overline{\sigma}\gtrsim5$) that are baryon-dominated on disc scales. These findings, in combination with the evidence for merger activity/interactions, reinforce that ordered rotation and merger-driven assembly are not mutually exclusive, as gas can rapidly resettle into a disc following an interaction. \\
\item Our resolved, kiloparsec-scale study of star formation relations demonstrates that, with current assumptions, including turbulence is crucial for accurately modeling star formation for dusty galaxies at high-redshift. While the data for ALESS3.1 and ALESS9.1 are offset from star formation relations that do not take turbulence into account, they align with gravo-turbulent models that include a multi-freefall correction factor which acknowledges that turbulence regulates the rate and timescale of star formation across different densities. Internal turbulence remains a critical regulator of star formation even in systems that appear to be dynamically cold rotating discs. 
\end{list}
Our study shows that the synergy between ALMA and JWST provides a holistic view of dusty star-forming galaxies in the early Universe and allows us to peer through the dust to witness the symphony of the star formation and gas dynamics that shape them.

\section*{Acknowledgements}

B.A.W., J.A.H., L.A.B. and C.-L.L acknowledge support from the ERC Consolidator Grant 101088676 (“VOYAJ”). P.S. is supported by the Leiden University
Oort Fellowship and the International Astronomical Union – Gruber
Foundation (TGF) Fellowship.  P.M.P. is funded by the Dutch Research
Council (NWO) through the Veni grant VI.Veni.222.364. M.R. is supported by the NWO Veni project “Under the lens” (VI.Veni.202.225). E.d.C., J.L. and A.B. acknowledge support from the Australian Research Council (project DP240100589). L.A.B. acknowledges support from the Dutch
Research Council (NWO) under grant VI.Veni.242.055. C.-C.C. acknowledges support from the National Science and Technology Council of Taiwan (NSTC 111-2112-M-001-045-MY3 and 114-2628-M-001-006-MY4), as well as Academia Sinica through the Career Development Award (AS-CDA-112-M02). R.D. acknowledges support from the INAF GO 2022 grant “The birth of the giants: JWST sheds light on the buildup of quasars at cosmic dawn” and by the PRIN MUR “2022935STW,” RFF M4.C2.1.1, CUP J53D23001570006 and C53D23000950006. T.R.G. acknowledge funding from the Cosmic Dawn Center (DAWN), funded by the Danish National Research Foundation (DNRF) under grant DNRF140. K.K. acknowledges support from the Knut and Alice Wallenberg Foundation. A.M.S. and I.S. acknowledge STFC (ST/ X001075/1). This paper makes use of the following ALMA data:ADS/JAO.ALMA$\#$2012.1.00307.S,$\#$2016.1.00048.S, $\#$2017.1.01163.S,$\#$2019.1.00771.S,$\#$2021.1.00024S, $\#$2022.1.00955.S. ALMA is a partnership of ESO (representing its member states), NSF (USA) and NINS
(Japan), together with NRC (Canada), NSTC and ASIAA (Taiwan), and KASI (Republic
of Korea), in cooperation with the Republic of Chile. The Joint ALMA Observatory is
operated by ESO, AUI/NRAO and NAOJ. The authors acknowledge assistance from Allegro, the European ALMA Regional Center node in the Netherlands. 

\section*{Data Availability}

The ALMA data used for this work are available on the ALMA Science Archive at https://almascience.eso.org. The specific JWST observations used for this work can be accessed via doi:10.17909/e33v-ga73.

\bibliographystyle{mnras}
\bibliography{ALESSTurbulenceKeyArXiv} 



\appendix
\section{ALESS17.1: An SMG at $z=2.574$ behind a LIRG at $z=1.538$}
\label{sec:17}

\begin{figure*}
  \centering
  \includegraphics[scale=0.28]{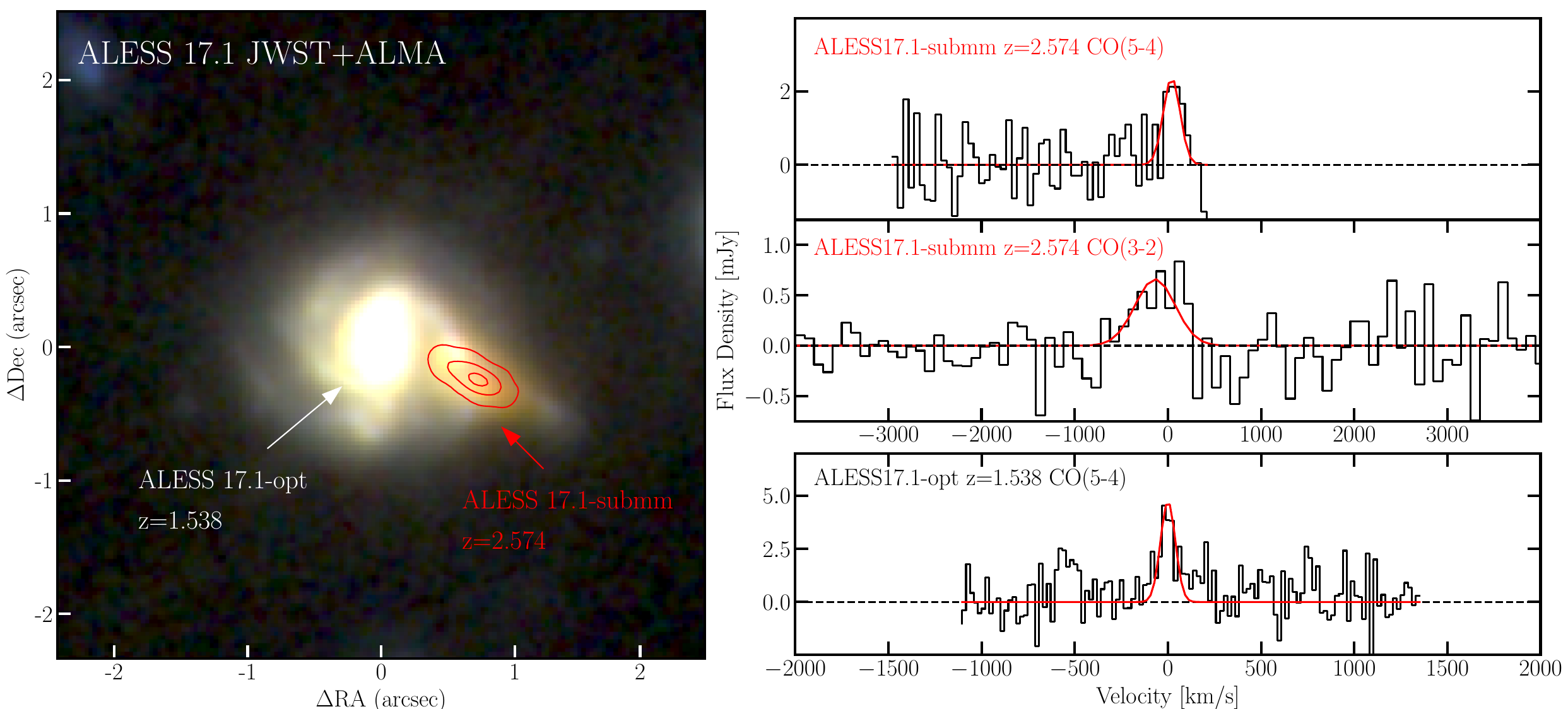}
  \caption{Left: NIRCam RGB image from \protect\cite{hodge_aless-jwst_2025} with positions of both ALESS17.1-submm indicated and 870$\mu$m contours from \protect\cite{hodge_alma_2019} at 5, 20 and 35 $\sigma_{\rm{rms}}$ Right-top/middle: Proposed archival ALMA CO line detections of redshifted CO(3--2) and CO(5--4) at $z=2.574\pm0.002$ for  ALESS17.1-submm (see also \protect\citealt{mckay_deep_2026}). This  is consistent with a JWST-based photometric redshift of $z\sim 2.57$ from \protect\cite{li_aless--jwst_2026}. Bottom: CO(5--4) emission line spectrum for ALESS17.1-opt.}
  \label{fig:ALESS17.1CO}
\end{figure*}

An additional source,  ALESS17.1, was targeted in the ALMA project  2022.1.00955.S [PI: Hodge] which is the basis of our study.  ALESS17.1 comprises a disturbed near-face-on, optically-bright spiral galaxy with faint submillimetre emission (which we will refer to as ALESS17.1-opt) and an apparently edge-on  submillimetre bright, but optically-undetected, source (ALESS17.1-submm) which intersects the spiral disk at a radius of $\sim$\,0.8$''$ or $\sim$\,7\,kpc (Fig.~\ref{fig:ALESS17.1CO}, \citealt[][]{hodge_kiloparsec-scale_2016,hodge_alma_2019}). ALESS17.1-opt is associated with a Chandra X-ray AGN \citep[log$L_{0.5-8 \rm{keV,corr}}=43.1\,\rm{erg\,s^{-1}},$][]{wang_alma_2013}.   

The bright submillimetre emission from this system was originally identified with ALESS17.1-opt on the basis of an ALMA position from the low-resolution Cycle~0 870-$\mu$m survey of \citet{hodge_alma_2013}, but subsequent higher resolution 870-$\mu$m ALMA observations showed the submillimetre emission was offset from ALESS17.1-opt \citep{hodge_kiloparsec-scale_2016}.  A spectroscopic redshift of $z=1.5397$ was obtained for ALESS17.1-opt  from the detection of redshifted H$\alpha$ emission by \citet{danielson_alma_2017},  supported by an earlier photometric redshift of $z=1.5^{+0.10}_{-0.07}$ from \citet{simpson_alma_2014}, with subsequent integral-field spectroscopy demonstrating that the H$\alpha$ emission extended across the location of ALESS17.1-submm \citep{chen_extended_2020}.   The apparent association of ALESS17.1-submm with ALESS17.1-opt was strengthened by the detection of CO(2--1) line emission at $z=1.538$ by  \citet{birkin_almanoema_2021} in $\sim$\,1$''$-resolution  ALMA Band 3 observations.

More recently,  JWST imaging of this system by \citet{hodge_aless-jwst_2025} identified a red stellar counterpart to the bright submillimetre source ALESS17.1-submm  which is undetected with the Hubble Space Telescope (HST). The very red F200W$-$F444W colour of ALESS17.1-submm and its brightness at 870\,$\mu$m are rare for a submillimetre galaxy at $z\sim1.5$, but could perhaps  be accounted for by an apparent edge-on orientation in an unusual orthogonal major merger and so these do not preclude the assignment of a redshift of $z\sim 1.54$ to  ALESS17.1-submm.  Hence this system was targeted as part of our study in CO(5--4) at $z\sim1.54$ using ALMA Band 6.  However, these high-resolution observations only showed a faint emission line at 227.06\,GHz  at the position of  ALESS17.1-opt corresponding to CO(5--4) at $z=1.538$ (Fig.~\ref{fig:ALESS17.1CO}).  This line has a flux of 0.68\,$\pm$\,0.11\,Jy\,km\,s$^{-1}$ and a FWHM = $210\pm40$\,km\,s$^{-1}$ (consistent with the CO(2--1) line width in \citealt{birkin_almanoema_2021}), implying a CO luminosity of $L'_{\rm CO(5-4)}=(35\pm6)\times10^8 \rm{\,K\,km\,s^{-1}}pc^2$.   The optical spectral and CO  properties of ALESS17.1-opt are consistent with those seen for typical massive dusty, star-forming galaxies at $z\sim 1.5$ \citep{birkin_almanoema_2021}.

The non-detection of redshifted CO(5--4) emission from ALESS17.1-submm is inconsistent with the submillimetre source lying at $z\sim1.5$.\footnote{Excess flux is seen at the position of ALESS17.1-submm in our ALMA Band 6 observations starting at 227.0\,GHz (the central frequency of the spectral window) and extending $\sim$\,1.0\,GHz to the edge of the spectral window with almost constant amplitude and a morphology similar to that seen for ALESS17.1-submm's dust continuum at 870\,$\mu$m and 1300\,$\mu$m.  Fitting a Gaussian profile to this feature assuming it is CO(5--4) yields a redshift of $z=1.546$ with a FWHM\,$\sim 1280\pm280\,\rm{km\,s^{-1}}$ at a nominal 7-$\sigma$ significance.  However, there is no corresponding broad line seen in the deep archival CO(2--1) observations at the position of ALESS17.1-submm  indicating that this broad feature is unlikely to be CO(5--4) emission.  We have checked for instrumental or calibration artefacts that might have caused this feature, but none were found  and we conclude that it is most likely  a transient instrumental artefact, although its underlying cause remains unknown.}   We therefore searched the ALMA archive for additional data that might help constrain the redshift of ALESS17.1-submm. We analysed shallow Band 3 and 4 observations  (120\,s and 90\,s integrations respectively) from project 2021.1.00024.S [PI: Bauer] which cover ALESS17.1 and identified two faint emission lines at 96.7974\,GHz and 161.2171\,GHz (Fig.~\ref{fig:ALESS17.1CO}) at the position of ALESS17.1-submm (see also \citealt{mckay_deep_2026}). The two lines correspond to redshift CO(3--2) and CO(5--4) emission and yield a redshift of  $z=2.574\pm 0.002$ for ALESS17.1-submm. The CO(5--4) line has a $\sim$\,7-$\sigma$ significance and a line flux of 0.54\,$\pm$\,0.12\,Jy\,km\,s$^{-1}$, but lies near the edge of the spectral coverage, while the CO(3--2) line was observed in two separate tunings and is detected at $\sim$\,3-$\sigma$ significance in each observation and 4.5\,$\sigma$ in the summed spectrum with a line flux of 0.36\,$\pm$\,0.07\,Jy\,km\,s$^{-1}$. This CO-based redshift is also supported by a JWST-based photometric redshift of $z\sim 2.57$ by \cite{li_aless--jwst_2026} for ALESS17.1-submm.

We conclude that the  redshift for the bright submillimetre galaxy ALESS17.1 is $z=2.574$, with the source lying in close proximity to a foreground LIRG-like galaxy (ALESS17.1-opt) at $z=1.538$. Further study of this source will be possible from the on-going ALMA project 2025.1.00761.S [PI: Westoby] observing the dust continuum in ALESS17.1 which will serendipitously  observe redshifted  [C{\sc i}] and CO(7--6) emission at $z=2.574$.

\section{Continuum Maps and ALESS3.1 CO(4--3) Maps}
\label{sec:map}

\begin{figure*}
    \centering
    \includegraphics[scale=0.47]{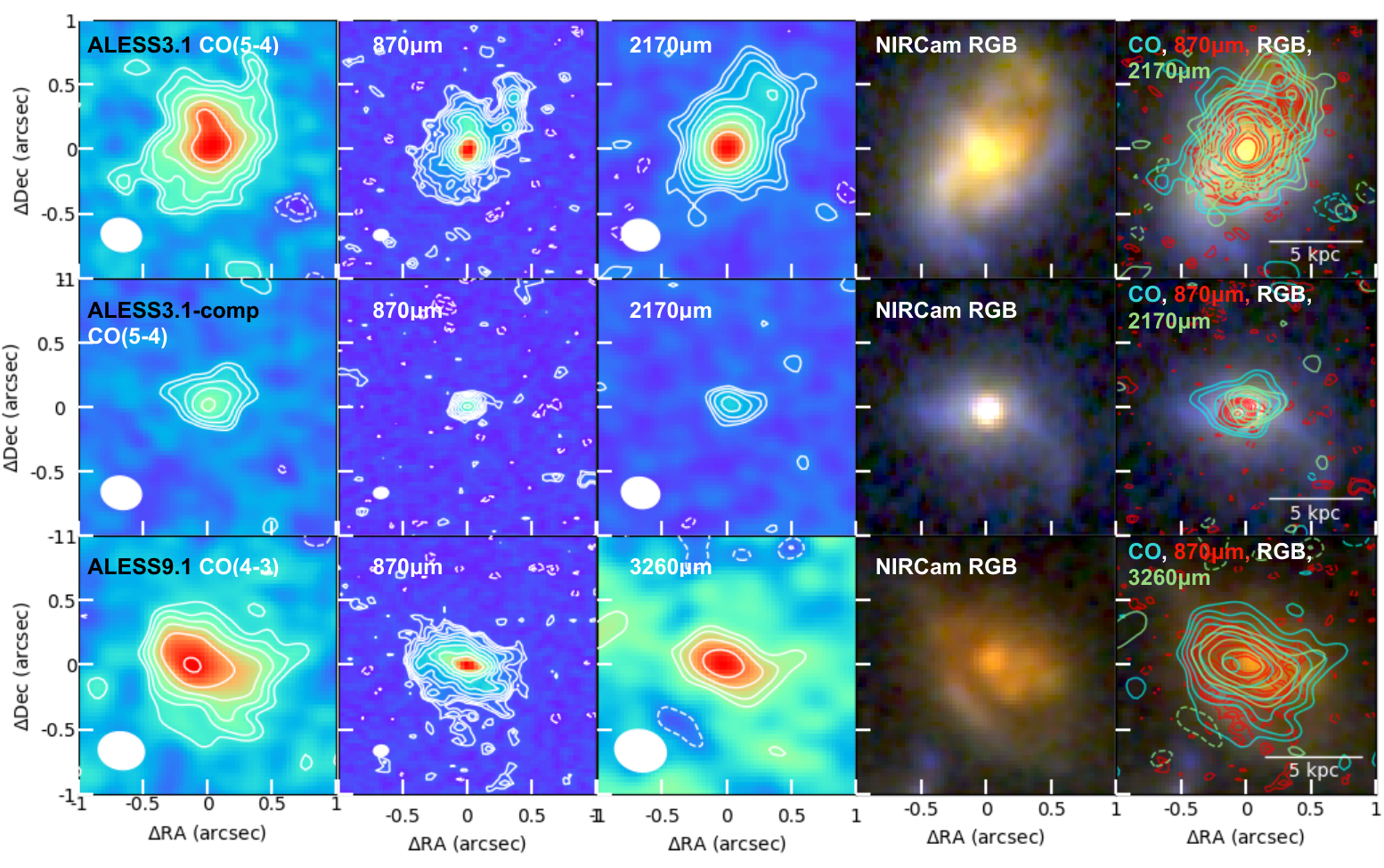}
    \caption{Multi-tracer imaging consisting of the CO moment zero maps (1st column), 870$\mu$m continuum (2nd column), continuum underlying the CO observations used in this work (3rd column: ALESS3.1 and ALESS3.1-comp 2170$\mu$m, ALESS9.1 3260$\mu$m) and the NIRCam RGB(F444W/F356W/F200W) image (4th column). The contours start at 2$\sigma_{\rm{rms}}$ and increase in powers of $\sqrt{2}$. The 5th column presents an overlay of all maps, with contours corresponding to the mid-$J$ CO cyan, 870$\mu$m continuum red and $\sim$2-3mm continuum green.}
    \label{fig:COcont}
\end{figure*}

The continuum maps for the three targets from the main text are shown in the second column of Fig \ref{fig:COcont}. 

\begin{figure}
    \centering
    \includegraphics[scale=0.7]{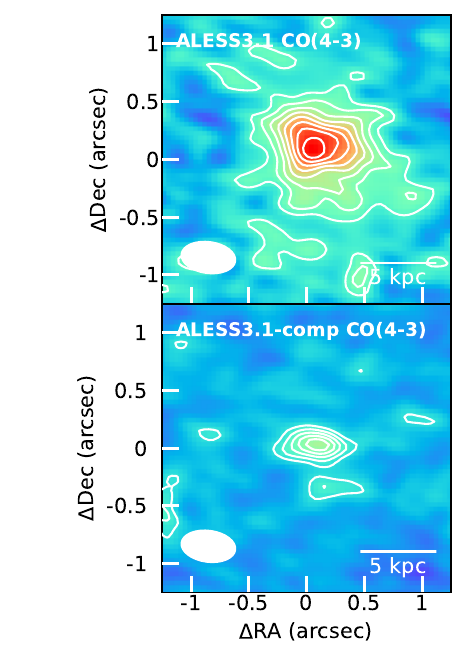}
    \caption{CO(4--3) emission for ALESS3.1 and ALESS3.1-comp. The beam is shown in the lower left corner and the contours start at 2$\sigma_{\rm{rms}}$ and increase in $\sigma_{\rm{rms}}$ steps.}
    \label{fig:ALESS3CO4to3}
\end{figure}

The lower-resolution ($\sim$4kpc) CO(4--3) maps for ALESS3.1 and ALESS3.1-comp are shown in Fig \ref{fig:ALESS3CO4to3} to give an impression of the extent and significance of the detections and the information about these maps is shown in Table \ref{tab:img2} alongside the information for the CO(5--4) maps for comparison. The CO(4--3) images have a beam size of 0.47$''\times$0.27$''$ and an RMS per channel of 230$\mu\rm{Jy\,beam^{-1}}$. 

\begin{table*}
\centering
\caption{Results from the naturally weighted imaging for both the CO(5-4) and CO(4-3) emission lines for ALESS3.1 and ALESS3.1-comp. Flux is measured using the same apertures on the zeroth moment maps, FWHM is calculated from Gaussian fitting of the spectra.}
\begin{tabular}{|l|c|c|c|c|c|c|}
\hline
Source ID & CO line & Band & Flux [$\rm{Jy\,km\,s^{-1}}$] & FWHM [$\rm{km\,s^{-1}}$] & SNR & $L'_{\rm{midJ-CO}}$ [$10^{11}\rm{\,K\,km\,s^{-1}}pc^2$]
\\ \hline
\multirow{2}{*}{ALESS3.1}
    & CO(5--4) & 4 & 1.62 $\pm$ 0.14 & 677 ± 68 & 11.45 & 0.33 $\pm$ 0.03  \\
    & CO(4--3) & 3 & 2.07 $\pm$ 0.19 & 776 ± 76 & 10.72 & 0.65 $\pm$ 0.06  \\ \hline
\multirow{2}{*}{ALESS3.1-comp}
    & CO(5--4) & 4 & 0.23 $\pm$ 0.05 & 187 ± 43 & 5.07 & 0.05 $\pm$ 0.01  \\
    & CO(4--3) & 3 & 0.22 $\pm$ 0.06 & 266 ± 85 & 3.78 & 0.07 $\pm$ 0.02  \\ \hline
\end{tabular}
\label{tab:img2}
\end{table*}

\section{Alternative Method - Dynamical Mass}
\label{sec:Mdyn}

In Section \ref{sec:KinematicModelling}, we find our estimates of the dynamical masses for ALESS3.1 and ALESS9.1 to be consistent within the uncertainties with the gas and stellar masses derived for the galaxies from the SED modelling \citep{li_aless--jwst_2026}. 

However, the estimates of the dynamical masses reported in this paper are subject to significant uncertainties arising from assumptions of the approach. For instance, the calculations assume that the galaxies have a single characteristic flat velocity (unlike the observed rotation curves for these galaxies), and that the stellar and gas discs are one single combined thin mass component.

To relax the above assumptions, we perform the exercise of comparing the circular speed profiles of our galaxies with those generated by the gravitational potential of our stellar and gas discs mass distributions. This alternative method is similar to the standard rotation curve decomposition approach at low-redshift, but note that we do not perform a full mass modelling. Instead, we simply compare the gravitational potential of our discs with that inferred from our kinematic observations.

For this, we follow \cite{mancerapina_impact_2022,mancera_pina_galaxy-halo_2025}, and use the software \textsc{galpynamics} \citep{iorio_off_2018}, which computes the potential and circular speed arising from our stellar and gas discs. For both components, we consider S\'ersic discs. The stellar disc is defined by the parameters reported for the GALFIT modelling of the F444W emission in \cite{hodge_aless-jwst_2025} and the gas disc is defined by the S\'ersic modelling in Section \ref{sec:mgm}. We also assume a realistic thickness of 500 pc, which is consistent with the thickness of local and high-redshift galaxies with similar mass and $V_{max}\,/\,\overline{\sigma}$ as our galaxies \citep{mancerapina_impact_2022,bacchini_3d_2024}. Finally, the mass of our gas disc is normalised to an $\alpha_{\rm{CO}}$ value of 1 $\rm{M_\odot}\, (K\,km\,s^{-1}\,pc^2)^{-1}$ .

\begin{figure}
    \centering
    \includegraphics[scale=0.42]{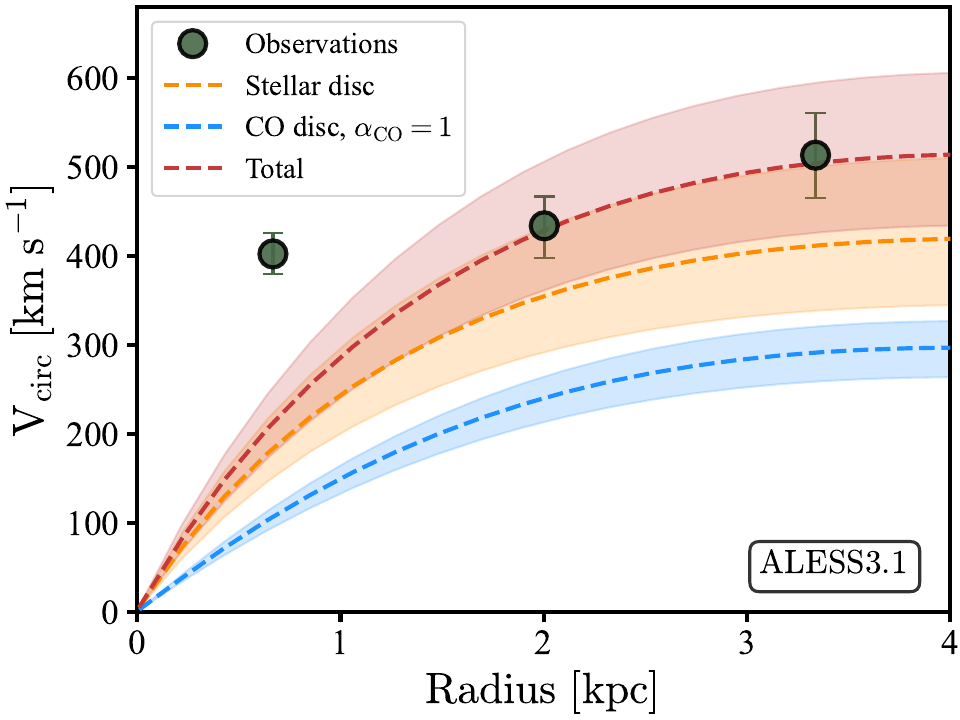}
    \includegraphics[scale=0.42]{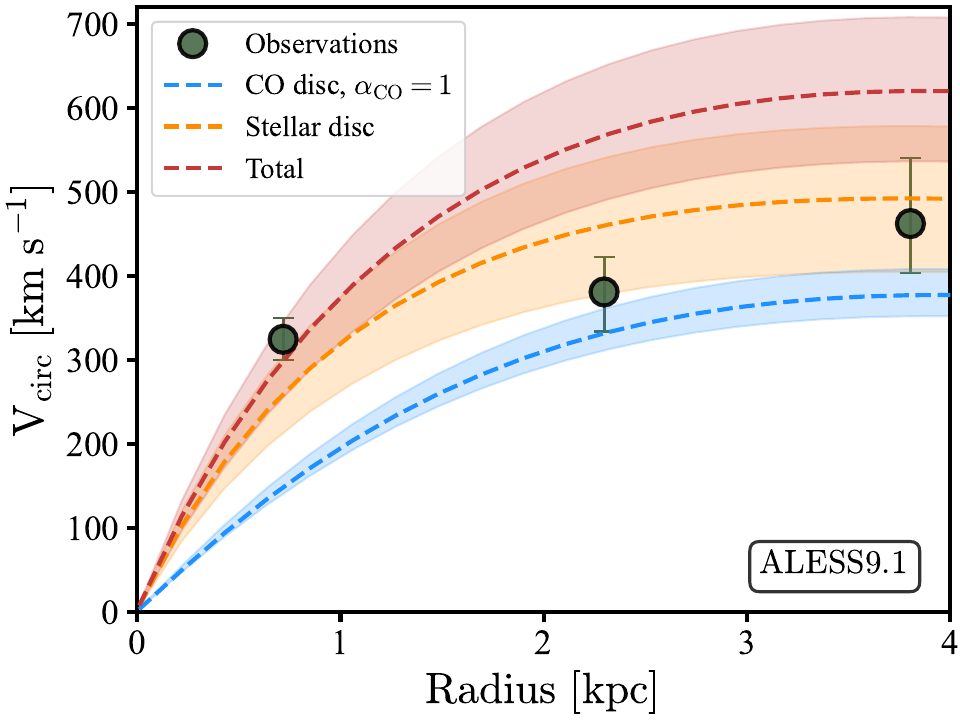}
    \caption{Pseudo-mass models for ALESS3.1 and ALESS9.1. The observed circular speed is shown with green markers and is compared against a total model (red curve) that included the circular speed components from the baryonic potential of the stars (orange curve) and gas (blue curve).}
    \label{fig:prc}
\end{figure}

Fig \ref{fig:prc} compares the circular speed provided by the baryons (stars in orange, gas disc in blue, total in red) with the observed circular speed of our galaxies (green markers). For ALESS3.1, we can see that the fiducial stellar mass and $\alpha_{\rm{CO}}=1\rm{M_\odot}\, (K\,km\,s^{-1}\,pc^2)^{-1}$ match nicely our observations, especially in the outer parts. The mismatch in the inner parts is potentially due to a bulge, which is implied for both ALESS3.1 and ALESS9.1 by the GALFIT residuals \citep[see Fig.5 in][]{hodge_aless-jwst_2025}. For ALESS9.1, the need for a low $\alpha_{\rm{CO}}$ value is much more dramatic, as the gravitational potential provided by the stellar disc is enough to explain the kinematics. We have performed additional tests (not shown), finding that if we reduce the stellar mass by three sigma, then there is room for $\alpha_{\rm{CO}}$=0.5 $\rm{M_\odot}\, (K\,km\,s^{-1}\,pc^2)^{-1}$ . While we do not perform a full rotation curve decomposition, our analysis shows clear dynamical evidence of low $\alpha_{\rm{CO}}$ values in our galaxies. This alternative method indicates $\alpha_{\rm{CO}}\lesssim1$, consistent with the estimated 2$\sigma$ limit of $\alpha_{\rm{CO}}\lesssim1.2$ in Section \ref{sec:KinematicModelling} and providing further support to our adoption of $\alpha_{\rm{CO}}=1\,\rm{M_\odot}\, (K\,km\,s^{-1}\,pc^2)^{-1}$ for the calculation of the total gas mass in Section \ref{sec:SpectraAndImaging} and in the resolved angular regions in Section \ref{sec:SFL}.

It is worth noting that the stellar masses derived from the integrated SED modelling (Table.\ref{tab:targetinfo}., \citealt{li_aless--jwst_2026}) are impacted by assumptions made about the IMF and SFH, presenting a key caveat to this alternative method. In particular, MAGPHYS, along with other SED codes, does not incorporate uncertainties from the choice of IMF into the uncertainties estimated for stellar masses, and accounting for IMF uncertainty would increase the errors in the mass-to-light ratio and therefore in the mass models. Additionally, the SFH parameterisation in MAGPHYS rises linearly at early ages and then declines exponentially. Random bursts of star formation are also superimposed onto the SFH. Therefore, the MAGPHYS model library accounts for a wide range of SFHs 
which is crucial to sample all possible mass-to-light ratios and so robustly constrain it \citep{da_cunha_alma_2015}. \cite{hainline_stellar_2011} and more recently \cite{jones_how_2026} find the chosen SFH of a model can have a significant effect on the stellar mass, although galaxies at higher redshifts are relatively insensitive to the SFH prior as the range of possible SFHs is reduced due to the shorter time since the galaxies' formation. The potential impact of these systematic uncertainties from the MAGPHYS choice of IMF and range of SFH on the stellar mass would further discourage exact determination of $\alpha_{\rm{CO}}$ from our dynamical mass estimates and should be considered when interpreting the pseudo-rotation curve method results.

\section{Alternative Method - Modified Black Body Star Formation Relations}
\label{sec:MBB}
\begin{figure*}
    \centering
    \includegraphics[width=\textwidth]{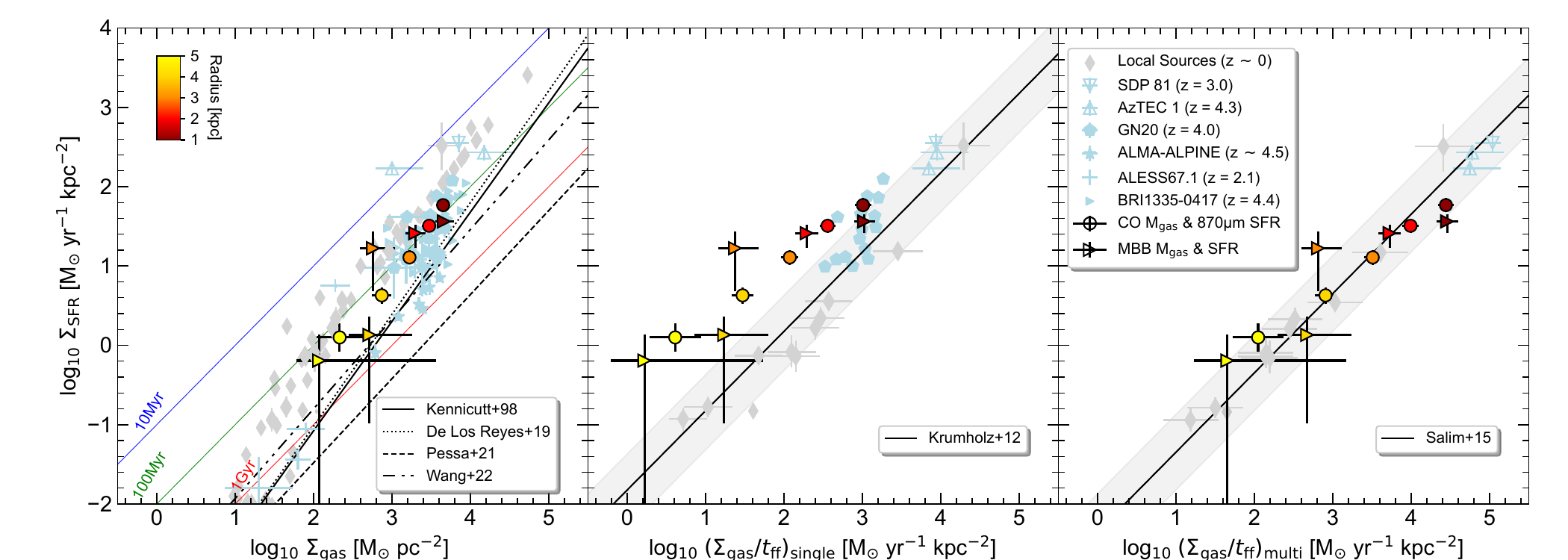}
    \caption{Star-formation relations for ALESS3.1, showing the MBB driven results (triangle), in comparison with the CO(5--4)-driven $\rm{M}_{\rm{gas}}$ and $870\mu$m-driven SFRs (circle, as in Fig. \ref{fig:KS}).}
    \label{fig:MBBKS}
\end{figure*}

Motivated by the proposed empirical link between CO(5--4) and dense gas \citep{daddi_co_2015}, we also estimate gas masses from dust masses derived via modified black body modelling of high-resolution continuum ($870\rm{\mu m}$ and the continuum underlying the CO), adopting the SMG gas-to-dust-ratio \citep[$\delta_{\rm{gdr}}/\alpha_{\rm{CO}}=63\pm7$,][]{birkin_almanoema_2021}. We also estimate star-formation rate from the modified black body modelling and use the conversion factor from \cite{murphy_calibrating_2011}. While the dust temperature of ALESS9.1 is unconstrained due to insufficient data\footnote{The continuum underlying the CO(4--3) line in ALESS9.1 (3260$\mu$m) lies further into the Rayleigh-Jeans tail than that of ALESS3.1 (2170$\mu$m) and the significance of the continuum detection is lower (as seen in Fig \ref{fig:COcont})}, the inferred gas surface densities and star formation rates from ALESS3.1 are consistent with the CO-based estimates (Fig  \ref{fig:MBBKS}).

\label{lastpage}
\end{document}